\documentclass[a4paper,11pt]{article}
\usepackage{jinstpub} 
\usepackage{lineno}
\usepackage{bm}
\usepackage{mathtools}
\usepackage{comment}

\newcommand{\strutlike}{\rule{0pt}{1.1\normalbaselineskip}}

\title{\boldmath Goodness-of-fit for multi-distribution neutrino cross-section measurements with shared events}

\author[a]{F. Mart\'inez L\'opez,}
\author[b]{L. Cooper-Troendle,}
\author[c]{S. Gardiner,}
\author[d]{P. Green,}
\author[a]{and T. Mohayai}
\affiliation[a]{Indiana University, Bloomington, IN 47405, USA}
\affiliation[b]{University of Pittsburgh, Pittsburgh, PA 15260, USA}
\affiliation[c]{Fermi National Accelerator Laboratory, Batavia, IL 60510, USA}
\affiliation[d]{University of Oxford, Oxford, OX1 3RH, United Kingdom}

\emailAdd{frmart@iu.edu}

\abstract{When multiple differential cross-section measurements are extracted from a common event sample, the same events contribute simultaneously to multiple distributions. This event-sharing structure imposes exact linear constraints among the bin counts, reducing the effective dimensionality of the measurement below the total number of bins. Since the covariance matrix obeys these inter-distribution constraints, it contains a rank deficiency. While limited numerical precision may inadvertently restore invertibility, the $\chi^{2}$ contributions along constrained dimensions will be arbitrary, yielding unphysically inflated $\chi^{2}$ values in global goodness-of-fit tests. We present the range-projected $\chi^{2}$, a test statistic that restricts the goodness-of-fit test to the subspace carrying independent statistical information, yielding a $\chi^{2}$ with $N_\text{bins} - N_\text{null}$ degrees of freedom, where $N_\text{null}$ is the number of independent constraints. We show that this rank deficiency is a predictable consequence of the event-sharing structure, and that $N_\text{null}$ decomposes into a structural contribution determined a priori from the binning geometry and a kinematic contribution that depends on the phase-space occupancy. The method is validated with an analytical toy model and a simulated neutrino--argon cross-section measurement including unfolding and multi-source systematic uncertainties.}


\arxivnumber{1234.56789} 

\begin{document}
\maketitle
\flushbottom

\section{Introduction}

Modern neutrino cross-section literature is experiencing an increase of published measurements reporting multiple differential distributions extracted from the same selected event sample. In these analyses, the same events contribute to multiple reported distributions, leading to strong correlations between them. Examples include MicroBooNE measurements of cross sections for final states with and without protons \cite{MicroBooNE:2024zkh,MicroBooNE:2024zwf} and multi-differential QE-like cross section measurements in kinematic imbalance variables \cite{MicroBooNE:2023tzj,MicroBooNE:2023cmw}. Other relevant examples include the multiple final-state kinematic projections from T2K \cite{T2K:2018rnz,T2K:2021naz}, and double-differential inclusive results from NOvA \cite{NOvA:2021eqi,NOvA:2024zmr,NOvA:2024rov}. Separate but related efforts have produced simultaneous measurements across different nuclear targets \cite{T2K:2020jav,MINERvA:2023kuz,MINERvA:2022djk} and combined neutrino-antineutrino results \cite{T2K:2020sbd}, where correlations arise from shared systematic uncertainties rather than shared events.

When multiple distributions are extracted from a common event sample, the event-sharing structure imposes exact linear constraints among the bin counts. If every selected event populates one bin in each distribution, then the total number of events in each distribution is exactly identical. More generally, the pattern of shared bin occupancy across distributions determines a set of constraints that reduce the effective dimensionality of the measurement below the total bin count. These constraints are a property of the measurement binning itself, independent of any particular statistical treatment.

The blockwise unfolding framework of Ref. \cite{Gardiner:2024gdy} provides a rigorous method for computing the full cross-distribution covariance matrices in such measurements. In this approach, the bins are organized into blocks, with each block corresponding to one reported differential cross-section distribution. Each bin block is unfolded independently using any standard method such as iterative D'Agostini unfolding \cite{DAgostini:1994fjx} or Wiener-SVD \cite{Tang:2017rob}. Separately, the full covariance matrix, including correlations between bins from different distributions, is constructed in reconstructed space and propagated to truth variables using a global unfolding matrix that consists of the unfolding matrices from each block in a block-diagonal format. The statistical covariance between bins in different blocks is determined by the number of events they share. The constraints mentioned above manifest as a rank deficiency in the statistical and systematic covariance matrices that respect the same event sharing patterns, as certain directions in bin space carry identically zero variance. The number of these degenerate directions is predictable from the binning geometry.

A natural use of these combined measurements is computing global goodness-of-fit metrics, such as a $\chi^{2}$ statistic, that test a model against all distributions simultaneously. Evaluating each distribution separately discards the correlations between blocks, both statistical fluctuations and systematic uncertainties correlated across distributions; a global test uses them to sharpen the comparison. Such global metrics are essential for interaction model tuning efforts \cite{Wilkinson:2016wmz,MINERvA:2019kfr,GENIE:2022qrc}, where measurements of multiple kinematic distributions provide complementary sensitivity to different aspects of the underlying physics. The event-sharing constraints, however, fix the data exactly, and a forward-folded prediction only approximately, along these directions. The data--model difference there is therefore zero in principle, and a global $\chi^{2}$ should receive no contribution from them. In practice it does. Independently of how the covariance matrix is built, finite numerical precision keeps it from being exactly singular, so it can still be inverted, while any systematic treatment that breaks the sharing lifts the constrained directions to spurious small variances. The residual along these directions, left small but nonzero by imperfect forward-folding, is then divided by a variance that is itself zero or spuriously small, an ill-defined ratio that numerical noise renders arbitrarily large. The result inflates the $\chi^{2}$, and even if the offending contributions were removed by hand the test would be left with the wrong number of degrees of freedom unless the constraints are properly identified and excluded.

We present a method to correctly compute a global goodness-of-fit test statistic for multi-distribution measurements with shared events. The approach identifies the constrained directions from the bin-combination structure of the measurement and projects it onto the subspace orthogonal to the constrained directions; we refer to the resulting test statistic as the range-projected $\chi^2$. Our contribution is twofold: we show that the rank deficiency of the covariance is a predictable consequence of the shared-event structure, rather than an incidental numerical degeneracy, and we provide a prescription to compute the number of constrained directions, with a structural component fixed a priori by the binning geometry and a kinematic component set by the phase-space occupancy. The statistical foundation of the projected statistic is the standard treatment of quadratic forms with singular covariance matrices \cite{Rao:1971}, a connection we make explicit while contrasting our approach with related methods.

The paper is organized as follows. Section~\ref{sec:constraints} describes the constraints imposed by the shared-event structure; Sec.~\ref{sec:test_statistic} constructs the range-projected test statistic and relates it to existing approaches; Sec.~\ref{sec:toy} and Sec.~\ref{sec:gen} validate the method on a toy model and a simulated measurement; and Sec.~\ref{sec:limitations} discusses the scope and limitations of the method.

\section{Constraints from shared events}
\label{sec:constraints}

Consider a neutrino cross-section analysis simultaneously measuring $B$ different flux-integrated differential cross section distributions, each with $N_b$ true bins. If the measurement is extracted from a common event sample then each selected event will populate exactly one bin in each distribution it belongs to. Recent MicroBooNE analyses \cite{MicroBooNE:2024yzp,MicroBooNE:2024zwf,MicroBooNE:2024pdj} have adopted the blockwise unfolding approach presented in Ref. \cite{Gardiner:2024gdy} that follows this approach to simultaneous cross-section measurements.

This method provides a systematic framework for reporting the correlations between the different extracted differential cross sections. It also gives a general prescription to construct statistical covariance matrices that respect the correlations between bins in different distributions or blocks. The full inter-block covariance matrix is first constructed in reconstructed space and then propagated through unfolding using a block-diagonal unfolding matrix formed from the direct sum of the individual unfolding matrices for each block. The result of the combined measurement is a vector of $N_{\text{bins}} = \sum_{b=1}^{B} N_b$ cross-section values with an $N_{\text{bins}} \times N_{\text{bins}}$ covariance matrix. The total covariance decomposes as $C_{\text{total}} = C_{\text{stat}} + C_{\text{syst}}$, where the statistical covariance component, computed from the shared-event structure, encodes the constraints described in the introduction.

\subsection{Statistical covariance from shared events}
\label{subsec:statistical_cov}

Inter-block correlations in the blockwise unfolding framework can be understood by considering two overlapping bins $X$ and $Y$, which may be partitioned into three non-overlapping regions $u$, $v$, and $w$. The events in the intersection $X \cap Y$ are assigned to bin $v$, while bin $u$ contains the events in the relative complement of $Y$ in $X$ (i.e. $X \setminus Y$), and bin $w$ corresponds to the relative complement of $X$ in $Y$ (i.e. $Y \setminus X$). Denoting the number of predicted events in each bin as $n_{i}$, and assuming these follow independent Poisson distributions, the covariance between the counts in bins $X$ and $Y$ is given by \cite{Gardiner:2024gdy}
\begin{equation}
    \begin{split}
        \text{Cov}(n_{X}, n_{Y}) &= \text{Cov}(n_{u} + n_{v}, n_{v} + n_{w})\\
        &= \text{Cov}(n_{u}, n_{v}) + \text{Cov}(n_{u}, n_{w}) + \text{Cov}(n_{v}, n_{v}) + \text{Cov}(n_{v}, n_{w})\\
        &= \text{Var}(n_{v}),
    \end{split}
\end{equation}
as by definition bins $u$, $v$, and $w$ have no events in common. This implies that the covariance between any two bins can be estimated from the variance of the events they share.

Monte Carlo (MC) events do not in general count as unity. It is standard practice to reweight event generator predictions to an alternative or tuned interaction model rather than simulate an entirely new sample, which would be computationally prohibitive \cite{NOvA:2020rbg,MicroBooNE:2021ccs}. Each event therefore carries a weight, and the variance for a set of such events is estimated from the sum of their squared weights. This gives us a general prescription to calculate the elements of the statistical covariance matrix as
\begin{equation}\label{eq:cstat_elements}
    (C_{\text{stat}})_{ij} = \sum_{e \in (i \cap j)} w_{e}^{2},
\end{equation}
where $w_{e}$ is the weight of event $e$, and the sum runs over events belonging to the intersection of $i$ and $j$. In the case of measured data the weights are $1$, and the covariance element simplifies to the number of events shared between the bins $n_{ij}$.

Throughout, we restrict the discussion to measurements in which each block uses the same number and structure for the reconstructed and true bins (as is the case in Ref.~\cite{MicroBooNE:2024yzp}). In this case the unfolding matrix of each block is square and invertible, so after propagation through the block-diagonal unfolding the covariance matrix in truth space has the same rank as its reconstructed-space counterpart. The rank has a direct interpretation in terms of the number of distinct event-sharing patterns across distributions, which we discuss next.

\subsection{Combination matrix and null space}
\label{subsec:combination_matrix}

The previous discussion shows that the statistical covariance between any two bins is determined solely by the events they share. We now formalize this observation to derive an exact formula for the rank of $C_{\text{stat}}$, and therefore for the number of directions in bin space that carry zero statistical variance.

For each event $e$, we can define the indicator vector $\boldsymbol{\phi}_{e} \coloneqq (\phi_{e,i})_{i = 1}^{N_{\text{bins}}}$ such that $\phi_{e,i} = 1$ if $i \in \text{bins}(e)$ and $\phi_{e,i} = 0$ if $i \notin \text{bins}(e)$, where $\text{bins}(e)$ is the set of bins event $e$ populates across all distributions. The indicator vector records which bins an event populates across all distributions. For example, in a two-block measurement with 3 bins each, an event populating bin 2 of block 1 and bin 1 of block 2 has indicator vector $\boldsymbol{\phi}_{e} = (0,1,0,1,0,0)$. The statistical covariance matrix can then be re-written in terms of indicator vectors as
\begin{equation}\label{eq:cstat_outer}
    C_{\text{stat}} = \sum_{e} w_{e}^{2} \, \boldsymbol{\phi}_{e} \, \boldsymbol{\phi}_{e}^{\mathsf{T}}.
\end{equation}

This outer-product decomposition makes the rank structure of $C_{\text{stat}}$ clear. The rank of the statistical covariance matrix equals the dimension of $\mathrm{span}\{\boldsymbol{\phi}_e\}$, which depends only on the distinct indicator vectors present, not on their weights. 


Since all weights satisfy $w_e > 0$, no outer-product term in Eq.~\eqref{eq:cstat_outer} vanishes, and the column space of $C_{\text{stat}}$ is exactly $\mathrm{span}\{\boldsymbol{\phi}_e\}$. We define the bin-combination matrix $\Omega$ as the matrix whose rows are the unique indicator vectors among $\{\boldsymbol{\phi}_e\}$. Each row of $\Omega$ represents a distinct pattern of bin occupancy across the blocks. By construction the row space of $\Omega$ equals $\mathrm{span}\{\boldsymbol{\phi}_e\}$, and therefore
\begin{equation}\label{eq:rank_formula}
    \mathrm{rank} (C_{\text{stat}}) = \mathrm{rank} (\Omega).
\end{equation}
The number of degenerate directions is then
\begin{equation}\label{eq:nnull}
    N_{\text{null}} = N_{\text{bins}} - \mathrm{rank} (\Omega).
\end{equation}

Physically, each of the $N_{\text{null}}$ directions corresponds to a linear combination of bins along which the measurement has identically zero variance. Such a combination is fixed by the event-sharing structure rather than measured, so the measurement has no statistical power to constrain it. Importantly, both $\Omega$ and $N_{\text{null}}$ are directly computable from the MC event-to-bin mapping alone, before constructing any covariance matrix, confirming that the constraints are a property of the measurement rather than the statistical treatment. In practice the mapping is read off from the MC sample, so the resulting count reflects not only the binning geometry but also the regions of phase space that the sample populates. We separate these two contributions in Sec.~\ref{subsec:null_space}.

A vector $\mathbf{x}$ is in the null space of $C_{\text{stat}}$ if and only if, for every event $e$
\begin{equation}\label{eq:null_condition}
    \sum_{i \in \text{bins}(e)} x_{i} = 0,
\end{equation}
or equivalently $\Omega \mathbf{x} = 0$. Each unique indicator vector acts as a linear constraint on the null-space vectors, requiring that the components of $\mathbf{x}$ sum to zero over the bins populated by the different events. The null space directions can be divided into two physically distinct categories: structural constraints that follow from the block topology alone, and kinematic constraints that depend on which regions of phase space are populated.

\subsection{Determination of the structural null space}
\label{subsec:null_space}

The combination matrix $\Omega$ presented previously is constructed from MC events by determining which bins each event populates and collecting the unique indicator vectors. However, the indicator vectors that can appear are constrained by two distinct mechanisms. The first is purely structural: the geometry of the binning scheme forbids certain bin combinations. The second is kinematic: correlations in model predictions between the different variables render some combinations unreachable even though they are geometrically allowed. These constraints are imposed by the underlying physics effects, mainly the differential cross section and the detector acceptance. In the following, we show that the structural constraints can be determined a priori, without generating MC events, from the bin edges used.

Our approach is to identify the geometrically allowed bin combinations directly from the bin edges. The key is to subdivide the binning finely enough that each resulting region lies within a single bin of every block, so that the allowed combinations can be read off directly. The structural null space follows from these combinations, as we develop below.

The variables a measurement is binned over form the set $\{v_{1}, \ldots v_{D}\}$. Each block $b$ uses a subset of these variables, $\mathrm{vars}(b) \subseteq \{v_{1}, \ldots v_{D}\}$. Restricting the discussion to combinations of single- and double-differential measurements, we have $|\mathrm{vars}(b)| \in \{1,2\}$ for every $b$. For each variable $v$ used in a given block $b$, the binning scheme defines a set of bin edges $E_{v}^{(b)}$ that partition the range of $v$ into intervals. In the case of a two-dimensional block, the edges in the binning variable may be slice-dependent, but nevertheless these can be collected into a single set of edges per variable across slices. For each variable $v \in \bigcup_{b} \mathrm{vars}(b)$, we can define the refined edges
\begin{equation}
    \mathcal{E}_{v} = \bigcup_{b \, : \, v \in \mathrm{vars}(b)} E_{v}^{(b)},
\end{equation}
with $M_{v} = |\mathcal{E}_{v}| - 1$ denoting the number of refined bins for variable $v$. Since $\mathcal{E}_{v}$ contains all the bin edges from every block that uses that variable, each refined bin is contained entirely within one original bin of every such block.

The common refinement of all blocks is the Cartesian product of the refined bins across all variables, denoted by $\mathcal{R}$. Each cell $c \in \mathcal{R}$ specifies a refined bin index for every variable in the measurement. For every block $b$, there exists a unique mapping $\beta_{b}(c)$ between each cell $c$ and an original bin in said block. The elements of the indicator vector $\boldsymbol{\phi}_{c} \coloneqq (\phi_{c,i})_{i=1}^{N_{\text{bins}}}$ associated to cell $c$ can be written as
\begin{equation}
    \phi_{c,i} = \begin{cases}
        1 & \text{if } i = \beta_b(c) \text{ for any block } b, \\
        0 & \text{otherwise.}
    \end{cases}
\end{equation}
If block $b$ does not use variable $v$ then $\beta_{b}$ is independent of the refined bin index in $v$. Therefore, multiple cells may produce identical indicator vectors.

The structural combination matrix $\Omega_{\text{struct}}$ is defined as the matrix whose rows are the unique indicator vectors from the cells of the common refinement $\{\boldsymbol{\phi}_{c} \, : \, c \in \mathcal{R}\}$. By construction, every indicator vector that appears in the $\Omega$ matrix obtained from the MC events is also present in $\Omega_{\text{struct}}$, therefore
\begin{equation}\label{eq:hierarchy}
    \mathrm{rank}(\Omega) \leq \mathrm{rank}(\Omega_{\text{struct}}) \leq N_{\text{bins}}.
\end{equation}
This hierarchy motivates a decomposition of the total number of degenerate directions
\begin{equation}
    \begin{split}
        N_{\text{null}} &= N_{\text{bins}} - \mathrm{rank}(\Omega_{\text{struct}}) + \mathrm{rank}(\Omega_{\text{struct}}) - \mathrm{rank}(\Omega)\\
        &= N_{\text{null}}^{\text{struct}} + N_{\text{null}}^{\text{kinem}}.
    \end{split}    
\end{equation}
The first term, $N_{\text{null}}^{\text{struct}}$, counts null directions that arise purely from the binning geometry and is computable a priori from the bin edges alone. The second term, $N_{\text{null}}^{\text{kinem}}$, accounts for additional null directions introduced by phase-space restrictions that exclude geometrically allowed combinations and can only be determined from Monte Carlo. For example, in a measurement of muon momentum and scattering angle, the common refinement may include cells for high-momentum backward-going muons. These cells are allowed by the binning and therefore appear in $\Omega_{\text{struct}}$. If they are empty because of the interaction kinematics or detector acceptance, their indicator vectors do not appear in $\Omega$, producing additional kinematic null directions. A description of the algorithm used to compute $\Omega_{\text{struct}}$ is presented in Appendix \ref{app:algorithm}.

The presence of kinematic null directions indicates that certain regions of the combined bin space are empty, which may signal that the binning extends beyond the kinematic reach of the measurement or that it is finer than the available statistics can support. In such cases, merging or removing the offending bins may be preferable to relying on a projection to handle the resulting constraints. The structural null directions, by contrast, are an unavoidable consequence of measuring multiple distributions from shared events. While their number depends on the binning, as discussed below, they cannot be removed entirely by any choice of bin edges.


The structure of the null space of $\Omega_{\text{struct}}$ depends on how the bin edges for the different blocks overlap. In the simplest case, when blocks do not share variables, every combination of one bin per block is allowed. The resulting null space is $(B-1)$-dimensional, corresponding to the $B-1$ independent constraints that require all block totals to be equal. When blocks share one or more variables, the allowed combinations are restricted and the dimension of the null space can be different from $B-1$. The number of null directions depends on how shared variable bin edges are aligned between blocks. Aligned slice edges decouple the measurement into different sub-problems, adding additional sum rules and increasing $N_{\text{null}}^{\text{struct}}$. On the other hand, misaligned edges couple adjacent distributions, potentially reducing the number of constraints. In the general case, the interplay between shared variables, bin edge alignment, and block topology makes a closed-form expression impractical, but the computation of $\mathrm{rank}(\Omega_{\text{struct}})$ from the bin edge specifications is straightforward.

The rank deficiency and null space discussed so far are properties of the reconstructed-space covariance, and, under the restriction to matched reconstructed and true binning adopted, to the unfolded covariance as well. If a block were instead unfolded from a larger number of reconstructed bins to a smaller number of true bins, the rank of the unfolded covariance would be bounded by the number of true bins, and the null space of the unfolded measurement could in principle be empty even though the covariance matrix is rank deficient in reconstructed space. We do not consider this case further here.

The structural and kinematic components of the null space respond differently to systematic uncertainties. Whether these cause the rank deficiency of $C_{\text{stat}}$ to persist, partially resolve, or fully resolve in the total covariance $C_{\text{total}}$ depends on the nature of the systematic treatment, which we discuss next.

\subsection{Effect of systematic uncertainties}
\label{subsec:systematics}

Previously, we established that the event-sharing structure imposes $N_{\text{null}} = N_{\text{bins}} - \mathrm{rank}(\Omega)$ exact linear constraints on the bin counts, which manifest as a rank deficiency of $C_{\text{stat}}$. By default, these same constraints apply to the construction of covariance matrices to describe various systematic effects, causing the same rank deficiencies. However, there is one notable exception in the case of computing the cross section covariance matrix, which we will discuss later. For now, we will focus on the treatment of other, more straightfordward systematic effects.

A systematic variation $s$ produces a modified vector of expected events $\mathbf{n}_{s}$, whose departure from the central-value prediction can be expressed in terms of a shift vector
\begin{equation}
    \boldsymbol{\xi}_{s} = \mathbf{n}_{s} - \mathbf{n}_{\text{CV}}.
\end{equation}
The different contributions to $C_{\text{syst}}$ are constructed from the outer product of these shift vectors. A null direction $\mathbf{x}$ of $C_{\text{stat}}$ remains a null direction of $C_{\text{total}}$ if and only if $\mathbf{x} \cdot \boldsymbol{\xi}_{s} = 0$ for every systematic variation in the uncertainty budget. We will now demonstrate that for a typical systematic variation, the shift vectors have zero projections onto the structural and kinematic null spaces.

Consider first the simplest case, where a systematic $s$ acts by reweighting the reconstructed event counts directly. Each event $e$ is assigned a modified weight $w_{e}^{(s)}$ while preserving its bin assignments. The shift vector can then be written as
\begin{equation}
    \boldsymbol{\xi}_{s} = \sum_{e} (w_{e}^{(s)} - w_{e}) \, \boldsymbol{\phi}_{e}.
\end{equation}
For any null vector $\mathbf{x}$ which satisfies the null-space condition of Eq.~\eqref{eq:null_condition}, the projection of $\mathbf{x}$ onto the shift vector gives
\begin{equation}
    \begin{split}
        \mathbf{x} \cdot \boldsymbol{\xi}_{s} &= \sum_{e} (w_{e}^{(s)} - w_{e}) \, (\boldsymbol{\phi}_{e} \cdot \mathbf{x})\\
        &= \sum_{e} (w_{e}^{(s)} - w_{e}) \sum_{i=1}^{N_{\text{bins}}} \phi_{e,i}\, x_{i}\\
        &= 0,
    \end{split}
\end{equation}
since the inner sum vanishes for every event. Direct reweighting therefore preserves the entire null space, adding no rank to $C_{\text{total}}$. Flux uncertainties, which modify the neutrino energy spectrum through per-event weights, are the classic example. When cross-section model variations and hadronic reinteraction uncertainties are implemented as per-event reweighting, they behave identically.

Systematic uncertainties can also be evaluated by directly comparing reconstructed event counts from independent MC samples simulated with nominal and varied parameters. Detector systematic uncertainties, which usually require independent samples with modified simulation parameters that cannot be captured by event reweighting, are the typical example where this prescription is applied. In this case, the shift vector takes the form
\begin{equation}
    \boldsymbol{\xi}_{s} = \sum_{e'} w_{e'}^{(s)} \, \boldsymbol{\phi}_{e'}^{(s)} - \sum_{e} w_e \, \boldsymbol{\phi}_{e},
\end{equation}
where the first sum runs over the events in the varied-model sample and the second sum over the nominal sample. Since each event in both samples populates exactly one bin per block, their difference projects to zero onto the structural null directions by an analogous argument to that used for the direct reweighting. Kinematic null directions, however, can be lifted if the variation scatters events into reconstructed-bin combinations absent in the nominal sample. In that case, new indicator vectors appear that are not rows of $\Omega$, potentially breaking the corresponding constraints. Therefore, despite their different treatment, detector systematic uncertainties also yield the same structural null directions determined by the binning scheme.

In contrast to these methods is a qualitatively different approach centered on varying the response matrix, commonly used in the calculation of cross-section systematic uncertainties. In this approach, adopted as the standard prescription in MicroBooNE for neutrino interaction uncertainties \cite{Gardiner:2024gdy}, a modified response matrix $\Delta_{s}'$, mapping true bins to reconstructed bins under the given variation, is constructed for each model parameter of interest. This matrix is obtained either by reweighting the events used to build the nominal response matrix or from an independent MC sample with varied simulation parameters. In this case, the predicted reconstructed counts for systematic variation $s$ are
\begin{equation}
    \mathbf{n}_{s} = \Delta'_{s} \, \boldsymbol{\varphi}_{\text{CV}},
\end{equation}
where $\boldsymbol{\varphi}_{\text{CV}}$ is the central-value truth prediction. The shift vector is then
\begin{equation}
    \boldsymbol{\xi}_{s} = (\Delta_{s}' - \Delta) \, \boldsymbol{\varphi}_{\text{CV}}.
\end{equation}
Since $\Delta$ is block-diagonal, the difference $\Delta'_{s} - \Delta$ acts independently on each block, projecting the nominal truth vector through block-specific changes in efficiency and migration. The per-event coupling that makes direct reweighting respect all null directions is absent in this prescription. When the weight of an event changes, it affects simultaneously every block it populates; the response-matrix method, in contrast, processes each block independently through its own sub-matrix. The projection of $\boldsymbol{\xi}_{s}$ onto any null vector, whether structural or kinematic, therefore involves differences in how each block responds to the same truth-level prediction, and in general is nonzero. Response-matrix variations thus break both types of null directions, regardless of how $\Delta_{s}'$ is obtained.

In practice, whether or not the null space of $C_{\text{stat}}$ survives in $C_{\text{total}}$, including these directions in a $\chi^{2}$ test via standard matrix inversion produces inflated values with the wrong effective number of degrees of freedom. Along these directions the measurement carries no statistical information, yet the systematic model, in particular the cross section systematic, lifts their variance to a small nonzero value. The inverse covariance then weights them by the reciprocal of that small value. Any residual between data and prediction that projects onto these directions is amplified, producing a contribution to the $\chi^{2}$ far in excess of one per degree of freedom, and not based in any physical data-model difference. We make this mechanism precise in Sec.~\ref{subsec:toy_gof}, where the excess is shown to be a noncentrality concentrated on the lifted directions. A valid goodness-of-fit test must therefore be restricted to the subspace where the measurement carries genuine sensitivity, which is the role of the projection method developed in the following section.

\section{Range-projected \boldmath{$\chi^{2}$} test statistic}
\label{sec:test_statistic}

Before constructing the test statistic, we must establish why restricting it to the non-null subspace discards no physical information. Along a null direction $\mathbf{x}$ of $C_{\text{stat}}$ the event-sharing constraint $\Omega\mathbf{x} = 0$ holds identically. The same selected events enter the bin combinations on both sides, so the data value along $\mathbf{x}$ is fixed by event conservation rather than measured, and a truth-level prediction respects the same constraint. The measurement therefore has no statistical power along $\mathbf{x}$, and no physical model can produce a genuine difference there. Forward folding a prediction through the detector response matrix does not in general preserve the constraint exactly, so a folded prediction can acquire a small non-zero component along $\mathbf{x}$. However, this residual is an artifact of the imperfect forward-folding rather than a real disagreement, and carries no statistical meaning, as the covariance assigns the direction no genuine variance. Discarding these directions removes no sensitivity, only the ill-defined contributions identified in Sec.~\ref{subsec:systematics}. This holds exactly to the extent that the event-sharing structure is exact; when sharing is only approximate the null directions acquire small but nonzero variance, a regime we return to in Sec.~\ref{sec:limitations}. Under the exact-sharing assumption, then, it is safe to formulate the global goodness-of-fit test statistic in the subspace orthogonal to the constrained directions, which we now construct.

The constraints are most naturally identified in reconstructed space, where the combination matrix is constructed. The non-null subspace can be obtained decomposing $\Omega^{\mathsf{T}} \Omega$. This matrix is a square $N_{\text{bins}} \times N_{\text{bins}}$ matrix with the same column space and null space as $\Omega$ itself, making it a more compact representation of the same information. We can write its eigendecomposition as
\begin{equation}\label{eq:oto_svd}
    \Omega^{\mathsf{T}} \Omega = V \, S \, V^{\mathsf{T}},
\end{equation}
where $S = \mathrm{diag}(s_{1}, \ldots, s_{N_{\text{bins}}})$ contains the eigenvalues in decreasing order and $V$ is the orthogonal matrix of eigenvectors. Of these eigenvalues, $r = \mathrm{rank}(\Omega)$ are nonzero. Partitioning $V = [V_{r} \, | \, \bar{V}_{r}]$, with $V_{r}$ collecting the eigenvectors associated with the nonzero eigenvalues and $\bar{V}_{r}$ those with zero eigenvalues, the columns of $V_{r}$ span the subspace of reconstructed bin space that is not constrained by the event-sharing structure while those of $\bar{V}_{r}$ span the null space. The orthogonal projector onto this subspace is
\begin{equation}\label{eq:proj_reco}
    P_{\text{reco}} = V_{r} \, V_{r}^{\mathsf{T}}.
\end{equation}
Note that $P_{\text{reco}}$ is identical to the projector onto the range of $C_{\text{stat}}$ as defined in Eq.~\eqref{eq:cstat_outer}, since $C_{\text{stat}}$ and $\Omega^{\mathsf{T}} \Omega$ share the same column space by construction. The combination matrix formulation makes explicit the fact that the projector is determined by the measurement geometry.

Cross-section measurements are typically reported in unfolded space, and so it is natural to assume this is the space where goodness-of-fit tests should be evaluated. The blockwise unfolding matrix $U$, which is block-diagonal and invertible, maps reconstructed-space vectors onto unfolded space, which is truth space in the case of an unregularized measurement. The non-null basis vectors propagate as $W_{r} = U \, V_{r}$. Since $U$ is not orthogonal in general, the columns of $W_{r}$ are not orthonormal, and the projector onto their span takes the general form
\begin{equation}\label{eq:proj_true}
    P_{\text{true}} = W_{r} \, (W_{r}^{\mathsf{T}} \, W_{r})^{-1} \, W_{r}^{\mathsf{T}}.
\end{equation}
Since $U$ is invertible, $W_{r}$ has the same rank as $V_{r}$ and the number of constrained directions is preserved. However, as shown next, the ordering of projection and unfolding matters as the commutation relation $U \, P_{\text{reco}} = P_{\text{true}} \, U$ generally does not hold. For this reason we choose to define the test statistic in reconstructed space, where the constraints originate.

In practice, limited numerical precision causes the computation of eigenvalues of $\Omega^{\mathsf{T}} \Omega$ that should be zero to yield small nonzero values. In many cases, these can be identified by setting a threshold relative to the largest eigenvalue (e.g., $s_{i} / s_{1} < 10^{-10}$). So long as physical eigenvalues do not span too many orders of magnitude and remain sufficiently far from zero, a clear spectral gap of several orders of magnitude can be observed between physical eigenvalues and numerical noise, making the identification unambiguous.

\subsection{Test statistic definition}
\label{subsec:chi2_definition}

The goodness-of-fit test can be evaluated in reconstructed space, where the residual vector is given by $\boldsymbol{\delta} = \mathbf{d} - \Delta \, \boldsymbol{\varphi}_{\text{pred}}$, where $\mathbf{d}$ is the background-subtracted data, $\Delta$ is the detector response matrix, and $\boldsymbol{\varphi}_{\text{pred}}$ is a truth-level model prediction. The range-projected $\chi^{2}$ is defined as
\begin{equation}\label{eq:chi2_proj_reco}
    \chi^2_{\text{proj}} = \boldsymbol{\delta}^{\mathsf{T}} \, ( P_{\text{reco}} \, C_{\text{total}} \, P_{\text{reco}} )^{+} \, \boldsymbol{\delta},
\end{equation}
with the superscript $^+$ denoting the Moore--Penrose pseudoinverse \cite{Penrose:1955vy}.

The pseudoinverse can be avoided entirely by reducing the problem to $r$ dimensions. One can verify that the pseudoinverse of $P_{\text{reco}} \, C_{\text{total}} \, P_{\text{reco}}$ takes the
closed form
\begin{equation}\label{eq:pinv_reco}
    ( P_{\text{reco}} \, C_{\text{total}} \, P_{\text{reco}} )^{+} = V_r \, (V_r^{\mathsf{T}} \, C_{\text{total}} \, V_r)^{-1} \, V_r^{\mathsf{T}}.
\end{equation}
Substituting into Eq.~\eqref{eq:chi2_proj_reco} and defining the reduced residual $\tilde{\boldsymbol{\delta}} = V_r^{\mathsf{T}} \, \boldsymbol{\delta}$ and the reduced covariance matrix $\tilde{C}_{\text{total}} = V_r^{\mathsf{T}} \, C_{\text{total}} \, V_r$, the test statistic becomes
\begin{equation}\label{eq:chi2_reduced}
    \chi^2_{\text{proj}} = \tilde{\boldsymbol{\delta}}^{\mathsf{T}} \, \tilde{C}_{\text{total}}^{-1} \, \tilde{\boldsymbol{\delta}},
\end{equation}
which involves only a standard inversion of the $r \times r$ matrix $\tilde{C}_{\text{total}}$. Under the null hypothesis, $\chi^2_{\text{proj}}$ follows a $\chi^2$ distribution with $r = N_{\text{bins}} - N_{\text{null}}$ degrees of freedom.

Without the projection, the $\chi^2$ test statistic takes the same numerical value whether computed in reconstructed or unfolded space, as the bias in the measurement introduced by a regularized unfolding cancel identically when compared to model predictions that are left-multiplied by the regularization matrix $A_{C} \equiv U \, \Delta$ \cite{Tang:2017rob}. The introduction of the projector breaks this equivalence when the projection is performed after unfolding. In unfolded-space the residual is defined as $\mathbf{r} = \hat{\boldsymbol{\varphi}} - A_{C} \, \boldsymbol{\varphi}_{\text{pred}}$, where $\hat{\boldsymbol{\varphi}} = U \mathbf{d}$ is the unfolded background-subtracted data. Using the propagated basis $W_r = U \, V_r$, one can construct an analogous test statistic in unfolded space as
\begin{equation}\label{eq:chi2_proj_unfd}
    \chi^2_{\text{unf}} = \tilde{\mathbf{r}}^{\,\mathsf{T}} \, \tilde{V}_{\text{total}}^{-1} \, \tilde{\mathbf{r}},
\end{equation}
where $V_{\text{total}} = U \, C_{\text{total}} \, U^{\mathsf{T}}$ is the total covariance matrix in unfolded space, and $\tilde{\mathbf{r}} = W_r^{\mathsf{T}} \, \mathbf{r}$ and $\tilde{V}_{\text{total}} = W_r^{\mathsf{T}} \, V_{\text{total}} \, W_r$ are the reduced residual and reduced covariance matrix in unfolded space. This quadratic form is not numerically equal to Eq.~\eqref{eq:chi2_reduced}. Expressing it in terms of reconstructed-space quantities yields
\begin{equation}\label{eq:chi2_unfd_expanded}
    \chi^2_{\text{unf}} = \boldsymbol{\delta}^{\mathsf{T}} \, U^{\mathsf{T}} U \, V_r \, ( V_r^{\mathsf{T}} \, U^{\mathsf{T}} U \, C_{\text{total}} \, U^{\mathsf{T}} U \, V_r )^{-1} \, V_r^{\mathsf{T}} \, U^{\mathsf{T}} U \, \boldsymbol{\delta},
\end{equation}
which reduces to Eq.~\eqref{eq:chi2_reduced} only when $U^{\mathsf{T}} U = I$, i.e. when the unfolding matrix is orthogonal.

The $\chi^{2}$ equivalence is restored if the projection is applied before unfolding. Projecting the reconstructed-space residual and covariance first, then propagating through the unfolding matrix, gives
\begin{equation}\label{eq:chi2_proj_then_unf}
    \begin{split}
        (U \, P_{\text{reco}} \, \boldsymbol{\delta})^{\mathsf{T}} \, (U \, P_{\text{reco}} \, C_{\text{total}} \, P_{\text{reco}} \, U^{\mathsf{T}})^{+} \, (U \, P_{\text{reco}} \, \boldsymbol{\delta}) &= \tilde{\boldsymbol{\delta}}^{\mathsf{T}} \, W_{r}^{\mathsf{T}} \, (W_{r} \, \tilde{C}_{\text{total}} \, W_{r}^{\mathsf{T}})^{+} \, W_{r} \, \tilde{\boldsymbol{\delta}}\\
        &= \tilde{\boldsymbol{\delta}}^{\mathsf{T}} \, G \, (G \, \tilde{C}_{\text{total}} \, G)^{-1} \, G \, \tilde{\boldsymbol{\delta}}\\
        &= \chi^2_{\text{proj}},
    \end{split}
\end{equation}
where the symmetric matrix $G = W_{r}^{\mathsf{T}} \, W_{r}$ acts as the unfolding matrix of the reduced system, and cancels similarly to the unprojected case. In the approach of Eq.~\eqref{eq:chi2_unfd_expanded}, the full residual, including the null-space components, is first rotated by $U$ into directions that are no longer orthogonal to the range. The unfolded-space projector then picks up this null-space contamination. Projecting first in reconstructed space avoids this by removing the null directions before unfolding can mix them with the physical subspace.

For this reason, Eq.~\eqref{eq:chi2_reduced} is the recommended definition of the range-projected $\chi^2$.

\subsection{Statistical validity}
\label{subsec:validity}

Under the null hypothesis, the residual $\boldsymbol{\delta}$ is drawn from a multivariate Gaussian distribution with mean zero and covariance $C_{\text{total}}$. The reduced residual $\tilde{\boldsymbol{\delta}} = V_r^{\mathsf{T}} \, \boldsymbol{\delta}$ is a linear transformation of a Gaussian vector, therefore Gaussian itself with mean zero and covariance $\tilde{C}_{\text{total}} = V_r^{\mathsf{T}} \, C_{\text{total}} \, V_r$. Provided that $\tilde{C}_{\text{total}}$ is positive definite, standard theory of Gaussian quadratic forms guarantees that $\tilde{\boldsymbol{\delta}}^{\mathsf{T}} \tilde{C}_{\text{total}}^{-1} \tilde{\boldsymbol{\delta}} \sim \chi^2(r)$. The positive definiteness of $\tilde{C}_{\text{total}}$ requires $C_{\text{total}}$ to have nonzero variance along every direction in the range of the projector $P_{\text{reco}}$, which is generally satisfied. This can be verified by checking that all eigenvalues of $\tilde{C}_{\text{total}}$ are strictly positive.

The validity of the range-projected $\chi^{2}$ can also be understood from the general theory of quadratic forms of correlated normal variables. Theorem 9.2.2 of Ref. \cite{Rao:1971} establishes that if $\mathbf{Y} \sim \mathcal{N}(0, \Sigma)$ with $\mathrm{rank}(\Sigma) = k$, then $\mathbf{Y}^{\mathsf{T}} \Sigma^{+} \mathbf{Y} \sim \chi^2(k)$ for any choice of generalized inverse $\Sigma^{+}$. Assuming the null hypothesis, the projected residual $P_{\text{reco}} \boldsymbol{\delta}$ is normally distributed with covariance $P_{\text{reco}} C_{\text{total}} P_{\text{reco}}$, which has rank $r$ by construction. The formulation of Eq.~\eqref{eq:chi2_proj_reco} with the Moore--Penrose pseudoinverse is  a specific instance of this general result. The key distinction from purely numerical approaches to handling singular covariance matrices is that the rank is not determined by an eigenvalue threshold applied to the covariance, but is predicted exactly via the combination matrix $\Omega$ from the structure of the measurement.

\subsection{Relation to alternative approaches}
\label{subsec:alternatives}

As a statistical object, the range-projected $\chi^2$ is an instance of the standard treatment of quadratic forms with singular covariance matrices described in Sec.~\ref{subsec:validity}. It is a generalized inverse restricted to the range of the covariance, distributed as a $\chi^2$ with degrees of freedom equal to the rank \cite{Rao:1971}. Dropping zero-variance directions from a $\chi^2$ is common practice in fits with degenerate covariance matrices. The contribution of this work is not the statistic itself but the recognition that the rank deficiency arises specifically and predictably from the shared-event structure, together with a prescription to compute its dimension a priori from the binning geometry. We discuss here how this differs from related approaches.

The most direct alternative is to regularize the covariance matrix numerically, for instance by performing a singular value decomposition and discarding eigenvalues below a fixed threshold relative to the largest. This is reliable only when a clean gap separates the null eigenvalues from the physical ones, which a realistic measurement does not guarantee. The null eigenvalues are lifted to small but nonzero values by any covariance component that does not preserve the event sharing; in cross-section measurements typically the response-matrix treatment of the interaction uncertainties. These lifted directions need not be separated from the smallest physical eigenvalues by a clean gap, so a threshold must then be chosen to sit between them. Setting it too low retains the inflating directions discussed next in Sec.~\ref{subsec:toy_gof}, while setting it too high discards directions carrying real information. The combination-matrix approach avoids this choice. In this case the rank is fixed by $\Omega$ from the measurement geometry, independently of the systematics-contaminated spectrum of $C_{\text{total}}$. A threshold is still used to identify the null space, but it is applied to $\Omega^{\mathsf{T}}\Omega$, where the gap is clean by construction, rather than to $C_{\text{total}}$ where systematics have filled it in.

The projection is distinct from, and complementary to, the regularization performed in unfolding. Methods such as Wiener-SVD \cite{Tang:2017rob} address the ill-posedness of inverting the response matrix, producing the regularization matrix $A_C$ that is applied to truth-level predictions before comparison. They do not address the event-sharing rank deficiency, which is a property of the statistical covariance and survives into unfolded space regardless of the unfolding method. The two operate on different sources of degeneracy and are used together. The event generator study of Sec.~\ref{sec:gen} unfolds each block with Wiener-SVD and applies the range projection on top.

A more complete alternative is to extract the cross section directly from a binned maximum likelihood fit, in which the truth-level bins are free parameters of interest and the systematic uncertainties enter as nuisance parameters with prior constraints (see, e.g., Ref. \cite{T2K:2023qjb}). The likelihood is evaluated in reconstructed space, where the inter-distribution correlations are carried by the shared events and the common nuisance parameters of the forward model. Therefore, the likelihood method never forms an inverted covariance over the bin counts, so the constrained directions are absorbed automatically and the goodness of fit can be assessed without confronting the rank deficiency at all. This approach is also more robust to the imperfect-sharing scenarios discussed in Sec.~\ref{sec:limitations}, where the constraints are only approximate. Its requirements are correspondingly greater: it needs the full forward model and nuisance parameterization, which are generally available only within the originating group. In this regard, the range-projected $\chi^2$ requires far less information.

For a group producing a blockwise-unfolded measurement, the range-projected $\chi^2$ is a natural global goodness-of-fit statistic. It is computed directly from the covariance and response matrices already used in the extraction, and is evaluated in reconstructed space as recommended in Sec.~\ref{subsec:chi2_definition}. It applies equally to an already-published measurement, as performed by generator comparison and tuning frameworks \cite{Stowell:2016jfr}. A standard data release reporting the unfolded data, unfolded covariance, and the block-diagonal unfolding matrix is sufficient to recover the reconstructed-space data and covariance by inversion; the reconstructed-space test then additionally requires the detector response matrix to forward-fold a model prediction. Where the response matrix is not available, the projection can instead be performed in unfolded space using the regularization matrix $A_C$, which is an increasingly standard release product. This unfolded-space test is less exact, as the non-orthogonal unfolding mixes a residual null-space contamination into the physical subspace, but it still removes the bulk of the inflation. The two approaches are complementary: the likelihood fit is more general when the complete model is available, while the projection provides a lightweight, a priori test from published results.

\section{Toy example}
\label{sec:toy}

The formalism developed previously is now illustrated with a toy model that captures the essential features of a multi-distribution neutrino cross-section measurement: a two-dimensional underlying physics process, a simple detector model, and a multi-block binning scheme in which events are shared across distributions. The toy is designed to be minimal but non-trivial. A three-block structure produces a rank-deficient statistical covariance matrix, and a confidence-region coverage study demonstrates the consequences of ignoring or correctly treating the resulting null directions.

\subsection{Model and measurement setup}

The toy model mimics a muon neutrino charged-current quasi-elastic--like interaction, producing a muon characterized by its momentum $p_\mu$ and scattering angle $\theta_{\mu}$. The joint truth-level differential cross section is parametrized as
\begin{equation}
    f(p_\mu, \cos\theta_\mu) = p_\mu^{\,k} \, e^{-k\,p_\mu / p_0} \, e^{\,\kappa \cos\theta_\mu} \, (1 + \alpha \, p_\mu \cos\theta_\mu),
\end{equation}
where the four parameters $(k, p_{0}, \kappa, \alpha)$ control the shape of the momentum spectrum, the level of forward peaking, and the strength of the momentum--angle correlation, respectively. A central-value (CV) model with $(k, p_0, \kappa, \alpha) = (3.0, 0.2, 2.0, 0.3)$ is used to build
the response matrix and estimate the covariance, while a ``fake'' model with $(k, p_0, \kappa, \alpha) = (3.3, 0.2, 1.8, 0.35)$ serves as the true underlying distribution from which pseudo-data are drawn. Truth events are generated via rejection sampling over $p_\mu \in [0.0, 1.2] ~ \mathrm{GeV}/c$ and $\cos\theta_{\mu} \in [-1.0, 1.0]$.

A simple detector model introduces two different effects. Momentum is smeared
multiplicatively, $p_\text{reco} = p_\text{true}(1 + \delta_p)$ with $\delta_p \sim \mathcal{N}(0, \sigma_p)$ and $\sigma_p = 0.05$, while the angle is smeared additively, $\theta_\text{reco} = \theta_\text{true} + \delta_\theta$ with $\delta_\theta \sim \mathcal{N}(0, \sigma_\theta)$ and
$\sigma_\theta = 1^\circ$.  A sigmoid efficiency function $\varepsilon(p_\text{reco}) = [1 + e^{-(p_\text{reco} - p_\text{thr})/w}]^{-1}$ with threshold $p_\text{thr} = 0.10 ~ \mathrm{GeV}/c$ and width $w = 0.02  ~ \mathrm{GeV}/c$ models the turn-on of the event acceptance at low momentum.

The analysis uses three bin blocks in reconstructed space: a $p_{\mu}$ marginal (5 bins), a $\cos\theta_{\mu}$ marginal (5 bins), and a two-dimensional distribution sliced in $\cos\theta_{\mu}$ (3 slices, each with 4 bins in $p_{\mu}$), for a total of $N_{\text{bins}} = 22$. The bin edges of the 2D block do not align exactly with those of the 1D marginals, which is the usual situation in real analyses. A CV sample of $N_{\text{MC}} = 500{,}000$ truth events is generated, processed through the detector model, and used to construct the response matrix $\Delta$. Pseudo-data samples are generated by drawing $N_{\text{data}} = 5{,}000$ events from the fake model.

\begin{figure}[!t]
	\centering
	\includegraphics[width=1.00\linewidth]{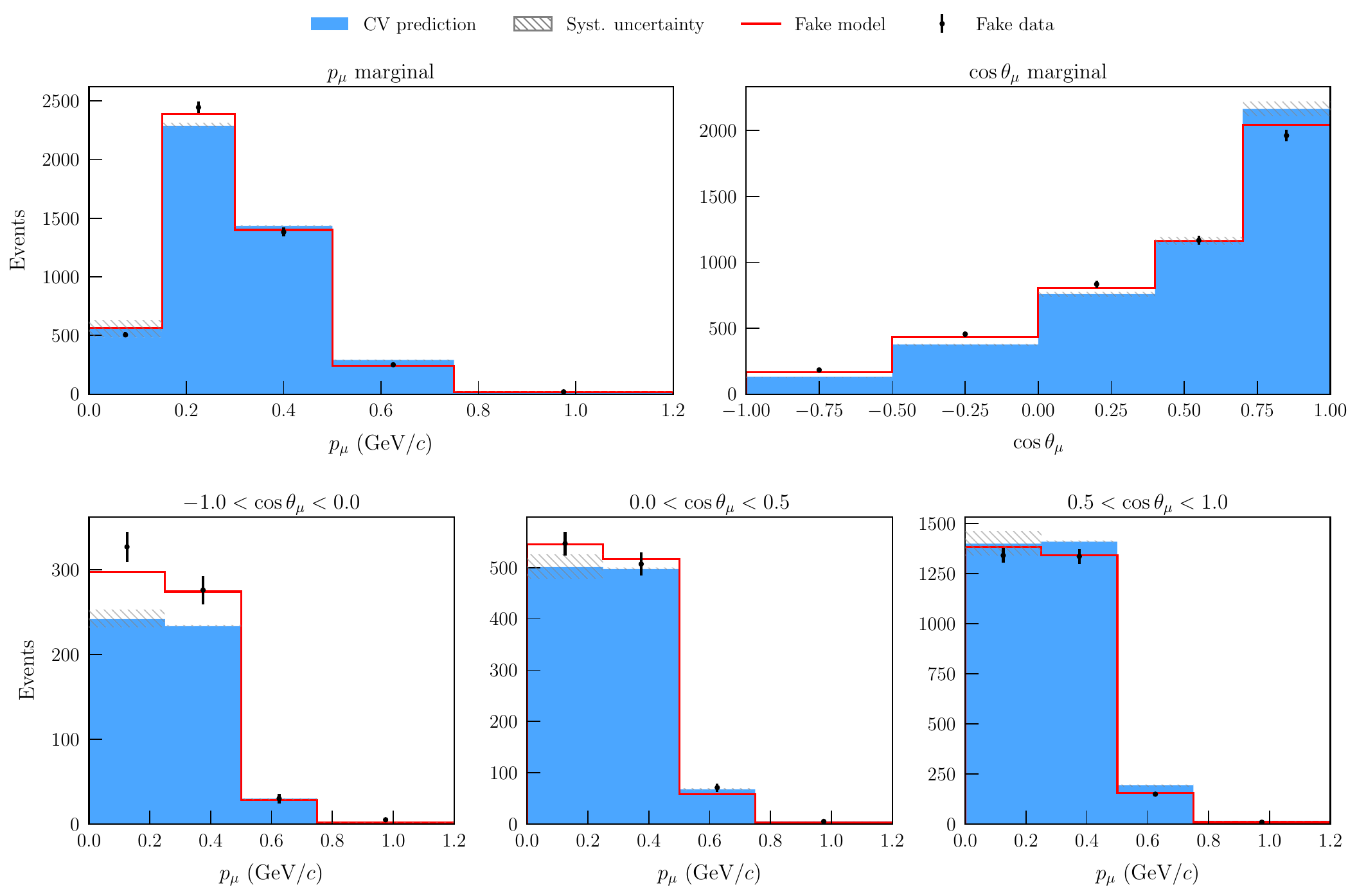}
	\caption{Predicted event distributions in the toy analysis blocks for the CV model (filled histograms) and the fake model (solid lines), together with a pseudo-data realization (points with statistical error bars). The shaded bands indicate the systematic uncertainty on the CV prediction. The top row shows the $p_\mu$ and $\cos\theta_\mu$ marginals; the bottom row shows the three $\cos\theta_\mu$ slices of the 2D block, each binned in $p_\mu$.}
	\label{fig:toy_distributions}
\end{figure}

Systematic uncertainties are estimated following the response-matrix approach described in Sec.~\ref{subsec:systematics}. Interaction-model variations are obtained by reweighting the CV sample to six alternative model configurations, each producing a varied response matrix $\Delta_{s}$. The shift vector $\boldsymbol{\xi}_{s} = (\Delta_{s} - \Delta) \, \boldsymbol{\varphi}$ captures the effect on the predicted reconstructed-space distribution. Detector variations are obtained by generating eight independent samples with perturbed detector parameters ($\sigma_p$, $\sigma_\theta$,
$p_\text{thr}$, and $w$), each producing a new response matrix from which a shift vector is computed similarly. The total systematic covariance matrix is $C_\text{syst} = N_\text{var}^{-1} \sum_s \boldsymbol{\xi}_{s} \, \boldsymbol{\xi}_{s}^\mathsf{T}$, scaled to data statistics. The statistical covariance is computed using Eq.~\eqref{eq:cstat_elements}, assuming unit event weights, which captures the inter-block correlations from event sharing.

Figure~\ref{fig:toy_distributions} shows the predicted event distributions for the CV and fake models in all three blocks of the toy analysis, together with a pseudo-data sample distribution and the systematic uncertainty bands.

\subsection{Null-space structure}

The structural combination matrix $\Omega_{\text{struct}}$ is constructed from the binning configuration of the three blocks, following the prescription from Sec.~\ref{subsec:null_space}. For this bin scheme, $\Omega_{\text{struct}}$ is rank $17$, yielding $5$ structural null directions. The combination matrix $\Omega$ computed from the MC event list also has rank $17$, confirming that all $5$ null directions are structural. No additional kinematic constraints arise for this choice of physics model and binning.

\begin{figure}[!t]
	\centering
	\includegraphics[width=0.80\linewidth]{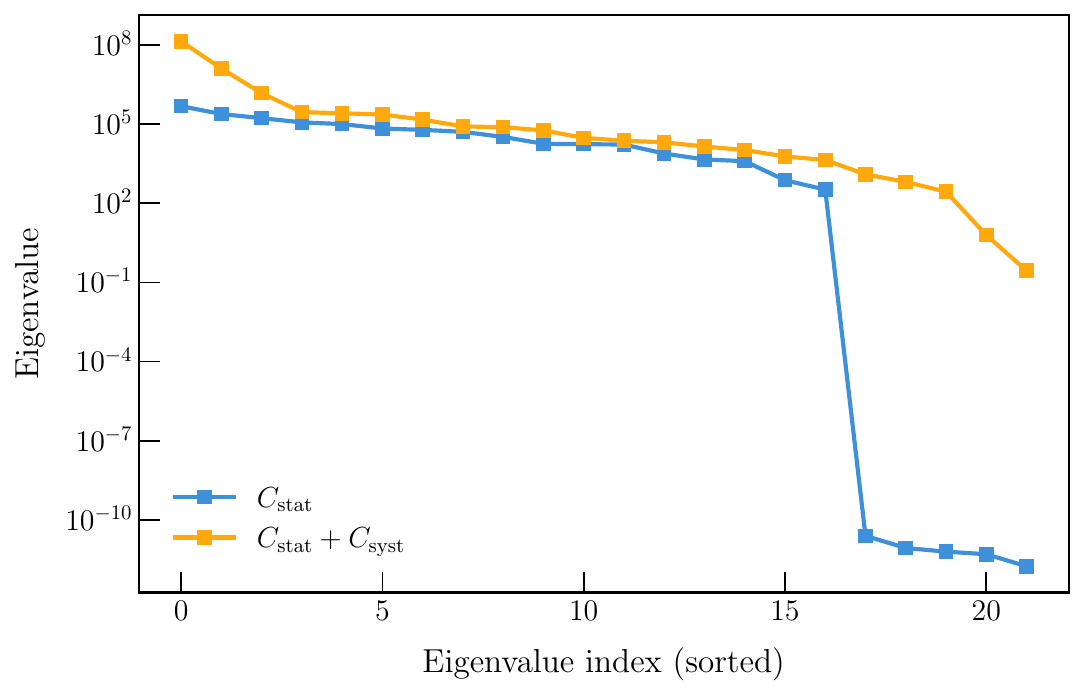}
	\caption{Eigenvalue spectra of the statistical covariance matrix $C_\text{stat}$ (blue circles) and the total covariance matrix $C_\text{total} = C_\text{stat} + C_\text{syst}$ (orange squares), for the toy analysis, sorted in descending order.}
	\label{fig:toy_covariance}
\end{figure}

The eigenvalue spectrum of $C_{\text{stat}}$ confirms this. It shows $5$ eigenvalues that are numerically zero, separated from the $17$ nonzero eigenvalues by a spectral gap of several orders of magnitude. Adding systematic uncertainties constructed via the response-matrix approach described in Sec.~\ref{subsec:systematics} lifts the $5$ null eigenvalues, as a consequence of the inconsistent folding through varied response matrices for each universe, making $C_\text{total} = C_\text{stat} + C_\text{syst}$ formally full rank. However, as shown in Fig.~\ref{fig:toy_covariance}, the lifted eigenvalues remain considerably smaller than the bulk of the spectrum. Including them in the test statistic introduces spurious contributions to the $\chi^{2}$, as demonstrated next. 

\subsection{Goodness-of-fit validation}
\label{subsec:toy_gof}

The fidelity of the projected and unprojected $\chi^{2}$ test statistics is assessed with a toy Monte Carlo exercise at the true hypothesis point. A high-statistics simulation of the fake model is used to estimate the event distribution on a fine reconstructed-space grid, defined by the union of all bin edges in $p_{\mu}$ and $\cos\theta_{\mu}$ across the three blocks, corresponding to the common refinement described in Sec.~\ref{subsec:null_space}. Each toy experiment draws a Poisson-fluctuated total event count and distributes them across the fine grid via multinomial sampling. These events are then projected onto the three analysis blocks, preserving the inter-block correlations from event sharing. Systematic fluctuations are added as correlated Gaussian throws from the Cholesky decomposition of $C_{\text{syst}}$. The $\chi^{2}$ is evaluated in reconstructed space, comparing each toy to the forward-folded prediction at the true hypothesis, for both the unprojected statistic, using the direct inverse of $C_{\text{total}}$, and the range-projected statistic of Eq.~\eqref{eq:chi2_proj_reco}.

\begin{figure}[!t]
	\centering
	\includegraphics[width=1.00\linewidth]{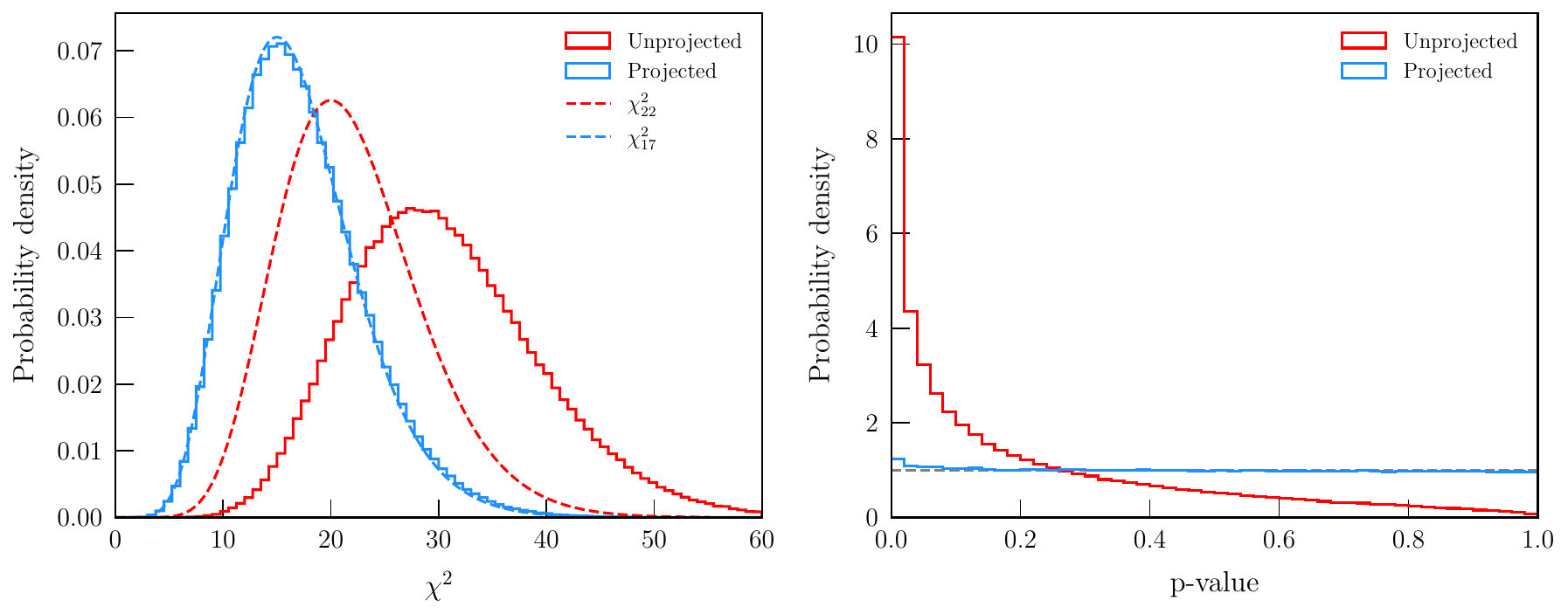}
	\caption{Goodness-of-fit validation for toy experiments generated at the true hypothesis. Left: distributions of the unprojected (red) and range-projected (blue) $\chi^2$ statistics, compared with the expected $\chi^2(22)$ and $\chi^2(17)$ distributions (dashed curves). Right: corresponding p-value distributions, with the false rejection rates at the $5\%$ significance level indicated.}
	\label{fig:toy_chi2_dist}
\end{figure}

Figure~\ref{fig:toy_chi2_dist} shows the results from $10^6$ toy experiments. The projected $\chi^{2}$ statistic closely follows the expected $\chi^{2}(17)$ distribution, with a mean of $17.23$ and a standard deviation of $6.01$, close to the expected values of $17$ and $\sqrt{2 \times 17} \approx 5.83$. In contrast, the unprojected statistic is shifted to significantly higher values with a mean of $30.59$, and a standard deviation of $8.96$, far above the expectation for a $\chi^{2}(22)$ distribution of $22$ and $\sqrt{2 \times 22} \approx 6.63$. The right panel shows the corresponding p-value distributions. The projected statistic yields an approximately uniform distribution, with a false rejection rate of $5.7\%$ at the $5\%$ significance level. The unprojected statistic produces a false rejection rate of $32.4\%$, more than six times the expected value. An analyzer using the unprojected statistic for goodness-of-fit testing would falsely reject the correct model roughly one third of the time.

\begin{figure}[!t]
	\centering
	\includegraphics[width=0.65\linewidth]{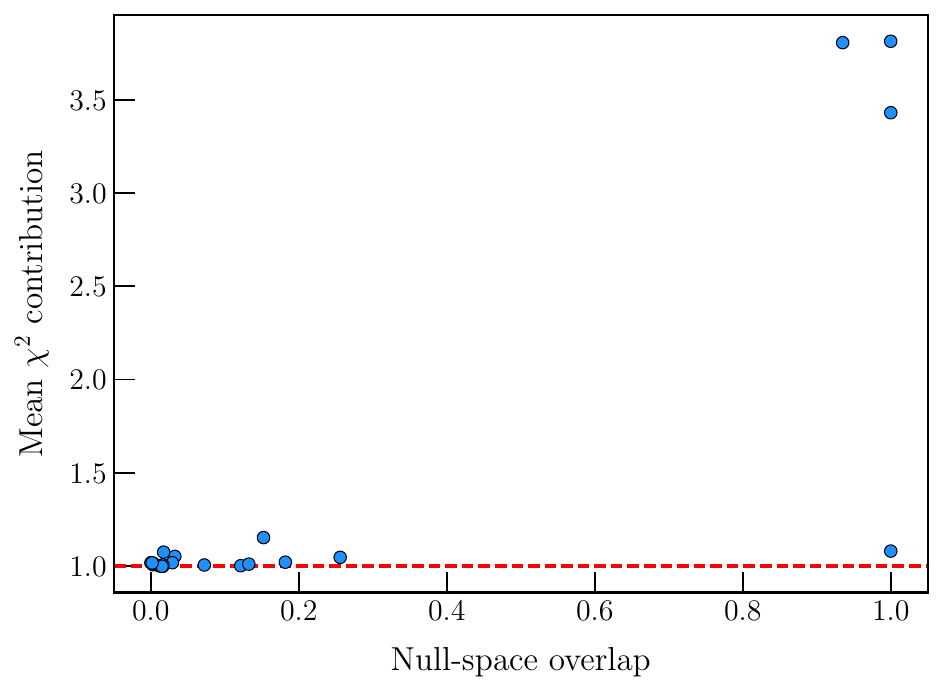}
	\caption{Mean contribution to the unprojected $\chi^{2}$ from each eigendirection of $C_{\text{total}}$, averaged over the toy ensemble, versus the null-space overlap $O_{j}$ from Eq.~\eqref{eq:overlap}.}
	\label{fig:toy_chi2_inflation}
\end{figure}

Writing the residual vector at the true hypothesis point as $\boldsymbol{\delta} = \boldsymbol{\delta}_{\text{rand}} + \mathbf{b}$, where $\boldsymbol{\delta}_{\text{rand}}$ has covariance $C_{\text{total}}$ and $\mathbf{b} \equiv E[\boldsymbol{\delta}]$ is the deterministic data--prediction residual (estimated as the mean over the toy throws), and diagonalizing $C_{\text{total}} = \sum_j \lambda_j \mathbf{v}_j \mathbf{v}_j^{\mathsf{T}}$, the unprojected statistic decomposes as
\begin{equation}
    \chi^{2}_{\text{unproj}} = \sum_j \frac{(\boldsymbol{\delta} \cdot \mathbf{v}_j)^{2}}{\lambda_j},
\end{equation}
with the expected value of each of these contributions given by
\begin{equation}
    \mathrm{E}\left[(\boldsymbol{\delta} \cdot \mathbf{v}_j)^{2} / \lambda_j\right] = 1 + \frac{(\mathbf{b} \cdot \mathbf{v}_j)^{2}}{\lambda_j}.
\end{equation}
Each eigendirection contributes as a noncentral $\chi^{2}(1)$ with noncentrality parameter $\psi_j = (\mathbf{b} \cdot \mathbf{v}_j)^{2} / \lambda_j$. On the directions overlapping the null space of $C_{\text{stat}}$ the eigenvalue $\lambda_j$ is the small lifted systematic variance, so any residual projecting onto them is amplified into a contribution far exceeding one.

To examine how the excess in the unprojected statistic is distributed across eigendirections, we quantify how much each eigenvector of $C_{\text{total}}$ overlaps the null space of $C_{\text{stat}}$. Using the projector onto the non-null subspace $P_{\text{reco}}$ from Eq.~\eqref{eq:proj_reco}, we define the overlap coefficient as
\begin{equation}\label{eq:overlap}
    O_{j} = \mathbf{v}_j^{\mathsf{T}} \, (I - P_{\text{reco}}) \, \mathbf{v}_j,
\end{equation}
which equals one for a direction lying entirely in the null space and zero for one in the range. Figure~\ref{fig:toy_chi2_inflation} shows the mean contributions to the $\chi^{2}$ test statistic across the toy ensemble against $O_{j}$. The $17$ non-null directions cluster around $O_j \approx 0$ with contributions consistent with unity. Across the ensemble the random part of the residual has the correct variance along every eigendirection, with the empirical variance matching $\lambda_j$ to better than $1\%$ on both the lifted and the range directions; the excess is therefore entirely noncentrality. The directions carrying the largest contributions are all those with substantial null-space overlap, where the small lifted eigenvalues amplify any residual projecting onto them. Summing over all directions, the noncentrality reproduces the observed excess, $\sum_j \psi_j \approx 8.6$, accounting for the unprojected mean of $30.6$.

\subsection{Parameter estimation}

\begin{figure}[!t]
	\centering
	\includegraphics[width=0.70\linewidth]{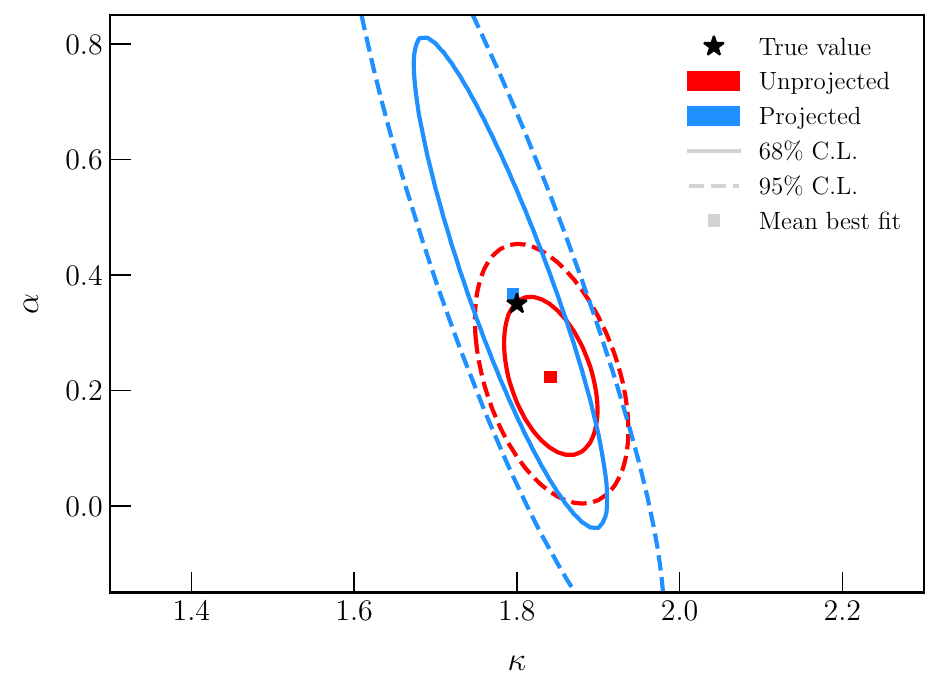}
	\caption{Confidence regions in the $(\kappa, \alpha)$ plane from the Asimov dataset, showing the 68\% and 95\% CL contours for the range-projected (blue) and unprojected (red) test statistics. The star marks the true parameter values; the squares indicate the mean best-fit points across $10^6$ toys.}
	\label{fig:toy_2d_scan}
\end{figure}

The impact of using the projected test statistic on parameter estimation is assessed with a two-dimensional confidence-region study. We scan the two parameters $(\kappa, \alpha)$ that control the angular shape and the momentum--angle correlation, holding the momentum-spectrum parameters $(k, p_0)$ fixed at their fake-model values, as these are well constrained by the $p_\mu$ marginal and largely decoupled from the null-space directions of interest. At each grid point the truth-level prediction is forward-folded through the nominal response matrix and compared to the pseudo-data, and the profiled test statistic $\Delta\chi^2 = \chi^2(\kappa, \alpha) - \chi^2_\text{min}$ is evaluated for both the unprojected and projected cases. Confidence regions are defined by the standard thresholds $\Delta\chi^2 < F^{-1}_{\chi^2_{n}}(\mathrm{CL})$ for $n$ scanned (or profiled) parameters, and their frequentist coverage is then measured directly from the toy ensemble.

Figure~\ref{fig:toy_2d_scan} shows the 68\% and 95\% confidence level contours from the Asimov dataset for both test statistics. The projected contours are correctly centered on the true parameter values, while the unprojected contours are shifted, with the 68\% region excluding the true point. The unprojected best fit values are biased, particularly in $\alpha$. The mean best-fit value across toys is $\hat{\alpha} = 0.22$, far from the true value of $0.35$, while the projected statistic recovers $\hat{\alpha} = 0.37$. The bias in $\kappa$ is milder ($\hat{\kappa} = 1.84$ for unprojected, $\hat{\kappa} = 1.80$ for projected, compared to the true $1.80$). This indicates that $\alpha$ has greater overlap with the null directions.

The frequentist coverage confirms these observations. At the 68.3\% confidence level, the projected statistic achieves nominal coverage in both the joint two-dimensional region ($68.2\%$) and the one-dimensional profiled intervals ($68.1\%$ for $\kappa$, $67.9\%$ for $\alpha$). The unprojected statistic suffers severe undercoverage: $36.5\%$ in the joint region, $44.6\%$ for $\kappa$, and $34.1\%$ for $\alpha$. The undercoverage is worst for $\alpha$, consistent with its stronger sensitivity to the null-space contamination. The projected contours are larger than the unprojected ones, reflecting the genuine sensitivity of the measurement once the ill-constrained directions are properly removed.

\section{Event generator study}
\label{sec:gen}

The toy study presented in the previous section validates the usage of the range-projected $\chi^{2}$ test statistic in a controlled setting, with an analytic cross-section model and known input parameters. Now, we demonstrate that the same pathology arises, and the same fix applies, when working with a realistic measurement setup. In this case, we use a realistic neutrino flux prediction, neutrino event generators, cross-section systematic uncertainties, and unfolding strategy. The latter is a key addition over the toy study, as it allows us to verify numerically the equivalence between the results in reconstructed and unfolded space, established in Sec.~\ref{subsec:chi2_definition}. In addition, we show that the projected test statistic recovers its discriminating power between generator predictions, correctly rejecting the wrong model while the unprojected statistic does not.

\subsection{Model and measurement setup}
\label{subsec:real_setup}

The central-value model is built from GENIE v3.6.2 \cite{Andreopoulos:2009rq} using the \texttt{AR23\_20i\_00\_000} tune on an $^{40}\mathrm{Ar}$ target. The fake-data model, which plays the role of the unknown true cross section, is NuWro v25.11 \cite{Golan:2012rfa} on the same target. Both generators are run with the SBND BNB $\nu_\mu$ flux over the energy range $0$--$10 ~ \mathrm{GeV}$. Since both predictions use the same flux, flux systematic uncertainties cancel in this comparison and are not considered; a measurement comparing against an external flux prediction would include them as an additional systematic.

Events are selected using a topological CC1$\mu$1$p$0$\pi$ definition applied identically to both generators. We require exactly one muon, exactly one proton above $40~\mathrm{MeV}$ kinetic energy, and no charged or neutral mesons above $50~\mathrm{MeV}$ in the final state. This topological selection captures CCQE, CCMEC, and CCRES events where the pion was absorbed in final-state interactions.

The detector model is the same as in the toy study. For each candidate event, we apply a Gaussian momentum and angular smearing ($\sigma_p = 5\%$, $\sigma_\theta = 1^\circ$), and a sigmoid efficiency threshold at $p_\text{thr} = 100~\mathrm{MeV}$ with a width $w = 20~\mathrm{MeV}$ to the primary muon.

The analysis binning follows the same three-block scheme as the toy study: a $p_\mu$ marginal, a $\cos\theta_\mu$ marginal, and a two-dimensional block sliced in $\cos\theta_\mu$. In the present sample, however, the highest-momentum bin of the most backward $\cos\theta_\mu$ slice of the two-dimensional block receives no selected events, as the CC1$\mu$1$p$0$\pi$ topology has negligible contributions from high-momentum backward-going muons. This empty bin is merged with its neighbor, leaving $5+5+(3+4+4)=21$ bins across the three blocks.

One million GENIE events are generated and used to construct the response matrix and estimate the systematic uncertainties. One million NuWro events are generated separately; $5{,}000$ of these are drawn as pseudo-data, with the remainder serving as the high-statistics truth prediction.

\subsection{Systematic uncertainties and unfolding}

Interaction-model systematic uncertainties are estimated using the GENIE Reweight framework \cite{Andreopoulos:2009rq}, which provides per-event weights at $\pm 1\sigma$ for each systematic knob. Fifteen knobs are varied, spanning the dominant sources of uncertainty for the CC1$\mu$1$p$0$\pi$ topology: four $z$-expansion coefficients for the CCQE axial form factor \cite{Meyer:2016oeg}, RPA screening and Pauli blocking corrections \cite{Nieves:2011pp}, the resonance axial and vector masses and overall normalization \cite{Rein:1980wg}, the CCMEC normalization, two non-resonant single-pion background parameters, and three final-state interaction parameters controlling the pion mean free path, absorption fraction, and charge-exchange fraction \cite{Dytman:2021ohr}. The interaction-model systematic covariance matrices are constructed via the response-matrix approach described in Sec.~\ref{subsec:systematics}, applying the varied $\Delta^{'}_{s}$ on the fixed CV prediction. Detector uncertainties are estimated as before, using $\pm 1\sigma$ variations of the four detector parameters ($\sigma_p$, $\sigma_\theta$, $p_\text{thr}$, $w$).

Each of the three distribution blocks is unfolded independently using the Wiener-SVD approach \cite{Tang:2017rob}, with the response matrix and total data covariance computed block-by-block. The per-block unfolding matrices are assembled into the block-diagonal matrix $U$, and the regularization matrix $A_C$ is also constructed from the per-block results. The full covariance is propagated through $U$ as $V_\text{total} = U \, C_\text{total} \, U^{\mathsf{T}}$, preserving the cross-block correlations from event sharing. Truth-level predictions are regularized with $A_C$ before comparison to the unfolded data.

\begin{figure}[!t]
    \includegraphics[width=1.00\linewidth]{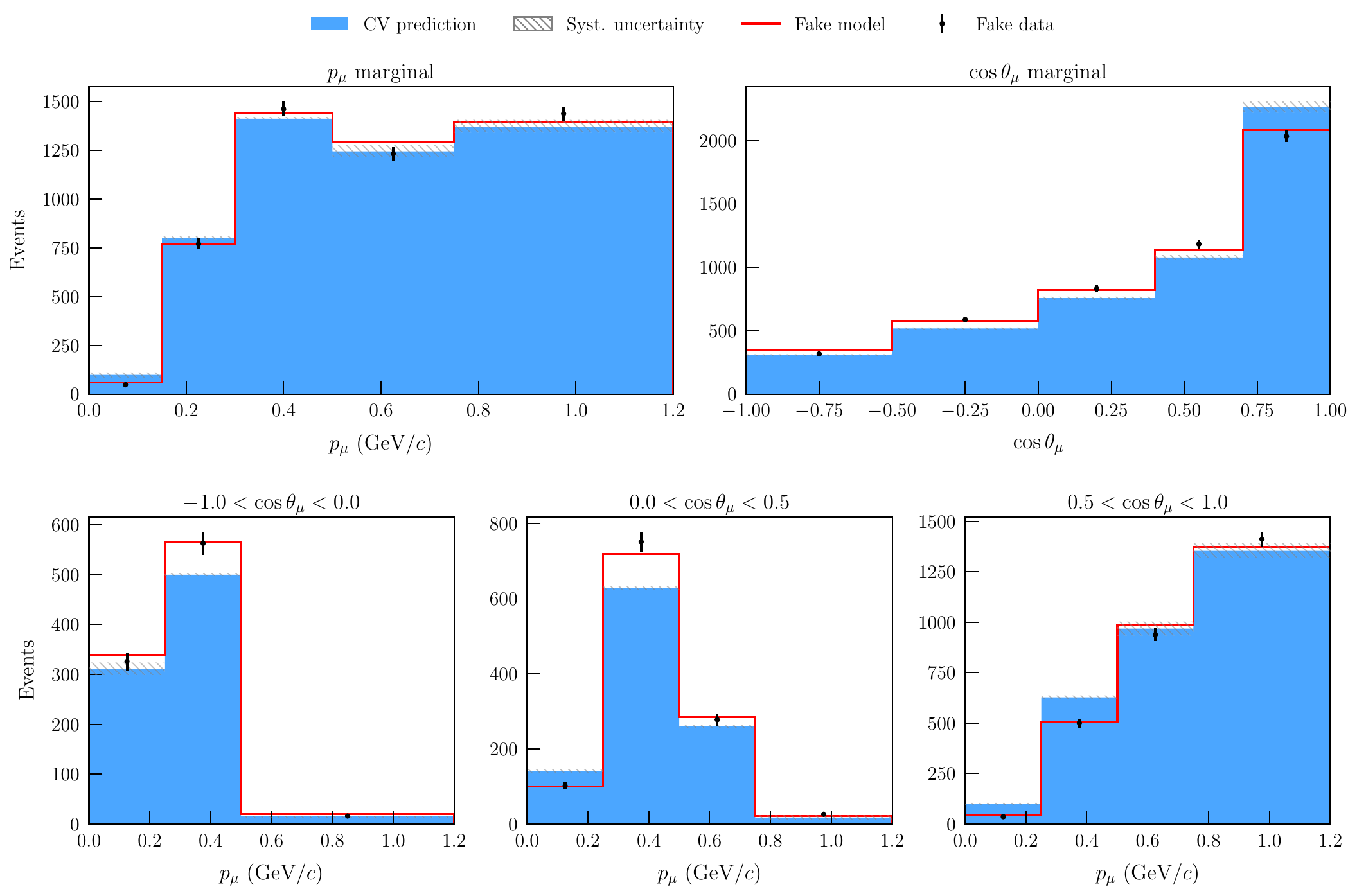}
    \caption{Reconstructed-space event distributions for the event generator study. The GENIE CV prediction (filled histograms) and systematic uncertainty bands (shaded) are compared to the NuWro prediction (red line) and a $5{,}000$-event NuWro pseudo-data realization (points with statistical error bars). The top row shows the $p_\mu$ and $\cos\theta_\mu$ marginals; the bottom row shows the three $\cos\theta_\mu$ slices of the 2D block.}
    \label{fig:real_distributions}
\end{figure}

\subsection{Results}

The structural combination matrix $\Omega_\text{struct}$, built from the three-block binning scheme, has rank $17$, yielding $N_{\text{null}}^{\text{struct}} = 21 - 17 = 4$ structural null directions. The full combination matrix $\Omega$, computed from the GENIE event list, has rank 16 and therefore reveals one kinematic null direction, bringing the total to $N_\text{null} = 5$. This is distinct from the bin merging mentioned in Sec.~\ref{subsec:real_setup}. In that case we merged a single bin of zero occupancy, whereas the kinematic null is a linear constraint among the populated bins. It arises because two cells of the common refinement corresponding to high-momentum backward-going muons are not populated by any selected events, a region where the CCQE-like cross section is strongly suppressed. The marginal bins themselves remain populated; only the fine-grained intersection of the high-momentum and backward-angle regions is empty. The projected test statistic therefore has $r = 16$ degrees of freedom.

Figure~\ref{fig:real_distributions} shows the reconstructed-space event distributions in all three blocks. The GENIE CV prediction and its systematic uncertainty band are compared to the NuWro fake-data realization and the NuWro prediction. The GENIE systematic band does not fully cover the NuWro prediction in several bins, reflecting genuine model differences that the $\chi^2$ test should be able to identify.

\begin{figure}[!t]
    \includegraphics[width=\textwidth]{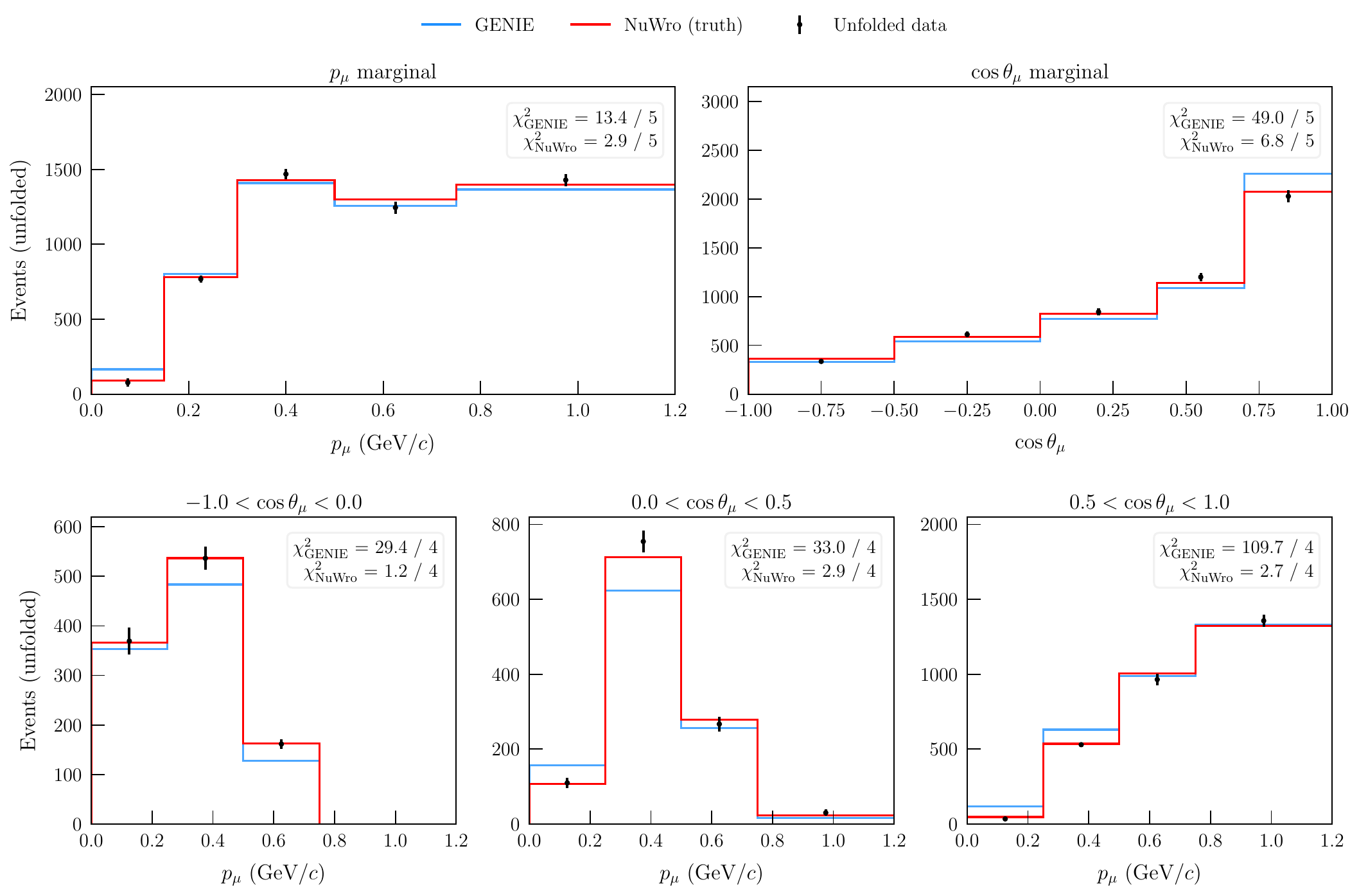}
    \caption{Unfolded event distributions for the event generator study. The GENIE (blue) and NuWro (red) truth-level predictions are regularized by the regularization matrix $A_C$ for comparison with the unfolded fake data (points with propagated uncertainties). Per-block $\chi^2$ values are annotated in each panel.}
    \label{fig:real_unfolded}
\end{figure}

Figure~\ref{fig:real_unfolded} shows the corresponding unfolded distributions, with the GENIE and NuWro truth-level predictions smeared through the regularization matrix $A_C$. Per-block $\chi^2$ and p-values, computed using only the block-diagonal covariance, are annotated in each panel. Comparisons between the unfolded NuWro fake data and the true NuWro model prediction yield p-values mostly above 0.32 and all above 0.05, confirming that the unfolding correctly recovers the true distribution within uncertainties at the per-block level

\begin{table}[!h]
	\caption{Global $\chi^2$ values for the NuWro fake data compared to the GENIE and NuWro truth-level predictions. The projected test statistic has $N_{\text{bins}} - N_{\text{null}} = 16$ degrees of freedom; the unprojected has $N_{\text{bins}} = 21$.}
	\begin{center}
        \begin{small}
			\begin{tabular}{ccc}
                \parbox{6em}{\centering Test statistic}         &  \parbox{4em}{\centering GENIE} & \parbox{4em}{\centering NuWro} \\[1mm] \hline
                \strutlike Unprojected $\chi^{2} / \text{ndof}$ &                      252.4 / 21 &                     121.9 / 21 \\[1mm]
                Projected $\chi^{2} / \text{ndof}$              &                      154.4 / 16 &                      15.7 / 16 \\[1mm]
            \end{tabular} 
        \end{small}
	\end{center}
	\label{tab:real_chi2}
\end{table}

Table~\ref{tab:real_chi2} summarizes the global $\chi^2$ results. The unprojected statistic yields $121.9 / 21$ for the NuWro prediction. This results in a disproportionate rejection of the correct model, entirely driven by the null-space contamination. The projected statistic gives $15.7 / 16$, consistent with the expected $\chi^2(16)$ distribution and confirming that the method is correctly calibrated. This was further validated with $10^5$ pseudo-experiments, which confirmed that the projected $\chi^2$ distribution follows $\chi^2(16)$ while the unprojected distribution does not follow $\chi^2(21)$, consistent with the toy study results of Sec.~\ref{subsec:toy_gof}.

The projected statistic retains full discriminating power against the wrong model. The comparison to GENIE yields $154.4 / 16$, a strong rejection indicating that the GENIE uncertainty budget cannot accommodate the NuWro prediction. Without the projection, both the correct and incorrect models are rejected, and the test provides no useful discrimination.

Finally, the projected $\chi^2$ was verified to give identical numerical values whether computed in reconstructed or unfolded space, for both the GENIE and NuWro predictions. This confirms the equivalence established in Sec.~\ref{subsec:chi2_definition} and validates the practical implementation of the projection in the presence of unfolding.

\section{Limitations and scope}
\label{sec:limitations}

The proposed method rests on the assumption that the event-sharing structure is exact. Every selected event populates exactly one bin in each distribution, so the constraints $\Omega \mathbf{x} = 0$ hold identically and the null space of $C_{\text{stat}}$, built according to Eq.~\eqref{eq:cstat_elements}, coincides with that of $\Omega$. Under this assumption the measurement has no statistical power along a null direction, so projecting it out discards no statistical information, as argued in Sec.~\ref{sec:test_statistic}. Therefore, our recommendation is that multi-distribution measurements be designed to preserve exact event sharing across all blocks, so the constrained directions can be projected out cleanly. The remainder of this section examines what happens when this requirement is relaxed, how imperfect sharing manifests, and how it might be handled.

In real measurements, this assumption is sometimes only approximate. Distributions extracted from a common sample may nonetheless use different phase-space limits, apply distribution-dependent background subtractions, or define their signal components differently, so that the populations entering different blocks are not strictly identical. Consider the simplest constraint that two block totals be equal, encoded by the direction $\mathbf{x}$ with entries $+1$ on the bins of one block and $-1$ on those of another. Its statistical variance is given by
\begin{equation}
    \mathbf{x}^{\mathsf{T}} C_{\text{stat}}\, \mathbf{x} = \sum_{e} w_e^2\,(\boldsymbol{\phi}_e \cdot \mathbf{x})^2 = \sum_{e\,\notin\,\text{both blocks}} w_e^2,
\end{equation}
which vanishes only when every event is shared. When a fraction $\epsilon$ of events is unshared, the variance is of order $\epsilon N$. In this case, the corresponding eigenvalue is lifted from zero into the body of the spectrum by the statistical model itself, before any systematic uncertainty is added. This is a qualitatively different situation from the inflation of Sec.~\ref{subsec:toy_gof}. There the variance along some directions was exactly zero in $C_{\text{stat}}$ and lifted only by the systematic model to a value unrelated to any statistical power, so dividing a residual by it produced an unphysical contribution. A direction carrying a genuine variance $\epsilon N$ from real event mismatch is not pathological, as the test weights it by its true variance and remains calibrated. The risk of imperfect sharing is therefore not renewed inflation but over-projection, i.e. discarding a direction that now carries a small but real amount of information.

The risk is confined to the a priori structural prescription, and it can be checked for. Three objects encode the null space under exact sharing, namely the combination matrices $\Omega_{\text{struct}}$ and $\Omega$ and the eigenspectrum of the statistical covariance matrix, and they respond differently when sharing is broken. The structural matrix $\Omega_{\text{struct}}$ depends only on the binning geometry and is blind to event occupancy entirely. When there is a genuine difference in phase-space acceptance between blocks, this can produce bin combinations absent from $\Omega_{\text{struct}}$, breaking the hierarchy of Eq.~\eqref{eq:hierarchy} so that $\mathrm{rank}(\Omega)$ exceeds $\mathrm{rank}(\Omega_{\text{struct}})$; the effect is visible in $\Omega$ but not in $\Omega_{\text{struct}}$. The empirical combination matrix $\Omega$ depends on which bin combinations are occupied, but not on how many events carry each, so it is blind to a violation caused only by missing events in certain blocks. In that case we can have $\mathrm{rank}(\Omega) = \mathrm{rank}(\Omega_{\text{struct}})$ even though the affected direction has acquired genuine variance. Only the spectrum of $C_{\text{stat}}$ reflects that variance, lifting the corresponding eigenvalue out of the null space.

\begin{figure}[!t]
	\centering
	\includegraphics[width=0.80\linewidth]{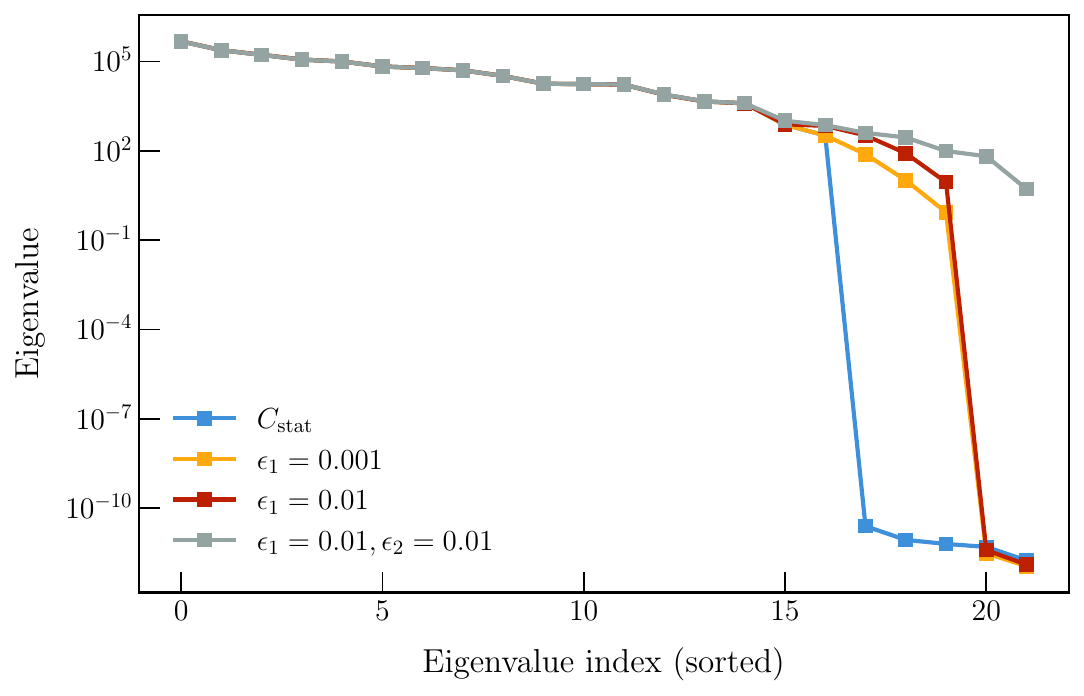}
	\caption{Eigenvalue spectra of the statistical covariance matrix $C_\text{stat}$ for the toy analysis, sorted in descending order, as events are removed at random from one or two blocks, breaking the exact event-sharing structure. Fractions $\epsilon_1$ and $\epsilon_2$ of events are dropped from the $p_\mu$ and $\cos\theta_\mu$ marginal blocks, respectively.}
	\label{fig:toy_covariance_spectra_epsilon}
\end{figure}

These two failure modes are illustrated with the toy model of Sec.~\ref{sec:toy}. The first is demonstrated by extending the $p_\mu$ acceptance of the two-dimensional block beyond that of the $p_\mu$ marginal, so that events with $p_\mu > 1.2~\mathrm{GeV}/c$ populate the former but not the latter. Of the $\mathcal{O}(5 \times 10^{5})$ generated events only seven fall in this region, yet their presence is enough to raise $\mathrm{rank}(\Omega)$ from $17$ to $18$, above $\mathrm{rank}(\Omega_{\text{struct}}) = 17$, and to make the affected block total fall short of the others by exactly those seven events. The rank signature is thus sensitive to even a single forbidden bin combination.

The second is demonstrated by removing a fraction of events from one block at random, so that the shared-event structure is broken without altering the binning. Figure~\ref{fig:toy_covariance_spectra_epsilon} shows the eigenvalue spectrum of $C_{\text{stat}}$ as this fraction is increased independently for blocks 1 ($p_{\mu}$ marginal) and 2 ($\cos\theta_{\mu}$ marginal). The formerly null eigenvalues, exactly zero under perfect sharing, are lifted into the bulk of the spectrum, filling the spectral gap that otherwise identifies the null space cleanly. The pattern set is unchanged, so $\mathrm{rank}(\Omega) = \mathrm{rank}(\Omega_{\text{struct}})$ throughout; only $C_{\text{stat}}$ reflects the violation.

These complementary sensitivities provide a practical set of consistency checks rather than a reason to abandon the geometric construction. Before relying on the structural projection, one should verify that $C_{\text{stat}}$ has the expected number of near-zero eigenvalues, separated from the bulk of the spectrum by a clean gap, that any difference between $\mathrm{rank}(\Omega)$ and $\mathrm{rank}(\Omega_{\text{struct}})$ is understood, and that the measured block totals are consistent within their correlated statistical uncertainties. The spectral-gap check assumes $C_{\text{stat}}$ is built from the sum of squared event weights following Eq.~\eqref{eq:cstat_elements}, which assigns exactly zero variance to the constrained directions. Estimators that quantify the statistical uncertainty differently may populate these directions with small nonzero values and need not show a clean gap even under exact sharing; there, the estimator-independent comparison between $\mathrm{rank}(\Omega)$ and $\mathrm{rank}(\Omega_{\text{struct}})$ is the reliable diagnostic. A deficit, $\mathrm{rank}(\Omega) < \mathrm{rank}(\Omega_{\text{struct}})$, is expected when allowed bin combinations are simply unpopulated; this is the kinematic null case seen in Sec.~\ref{sec:gen}, handled correctly by the empirical $\Omega$. An excess, $\mathrm{rank}(\Omega) > \mathrm{rank}(\Omega_{\text{struct}})$, instead signals that events occupy structurally forbidden combinations, the hallmark of broken sharing. When the diagnostics are consistent with exact sharing, the a priori count of Sec.~\ref{sec:constraints} applies as derived. When they are not, the structural null space over-counts the constrained directions, and projecting onto it would discard information; this signals that the binning or selection definitions should be reconciled across blocks, or that a treatment retaining the affected directions, such as marginalizing over nuisance parameters describing the differences (Sec.~\ref{subsec:alternatives}), is more appropriate.

\section{Summary and conclusions}
\label{sec:conclusions}

Multi-distribution cross-section measurements that share a common event sample contain exact linear constraints among their bin counts. When the statistical covariance accounts for the inter-distribution correlations, these constraints appear as a rank deficiency, and a global $\chi^{2}$ that ignores them is ill-defined. The null directions either prevent inversion or, once lifted by systematic uncertainties, produce unphysically inflated values. We have presented the range-projected $\chi^{2}$, a goodness-of-fit statistic that restricts the test to the subspace carrying independent statistical information, yielding a $\chi^{2}$ with $N_{\text{bins}} - N_{\text{null}}$ degrees of freedom.

This work makes two key contributions. First, we have shown that this rank deficiency is a predictable consequence of the shared-event structure rather than an incidental numerical degeneracy. The constrained directions are exactly those along which the data carry no statistical power, so removing them discards no physical information. Second, we have provided a prescription to compute the constrained directions a priori from the binning geometry, decomposing $N_{\text{null}}$ into a structural contribution fixed by the bin edges and a kinematic contribution set by the phase-space occupancy. Because the rank is determined from the measurement geometry rather than from the systematics-contaminated covariance spectrum, the method requires no eigenvalue threshold on the full covariance, where no clean separation between null and physical directions needs to exist.

The method was validated in two settings. In an analytical toy model the projected statistic follows the expected $\chi^{2}(17)$ distribution, while the unprojected statistic is inflated to a mean of $30.6$ and over-rejects the correct model in roughly one third of experiments; we traced the inflation to a noncentrality on the directions lifted out of the null space by systematic uncertainties. In a simulated neutrino--argon measurement with unfolding and multi-source systematic uncertainties, the projected statistic gives $15.7/16$ for the correct model and $154.4/16$ for an incorrect one, recovering both calibration and discriminating power, whereas the unprojected statistic rejects both models. The two studies share five null directions of differing composition, structural in the toy and a mixture of structural and kinematic in the simulated measurement. We note that the projected statistic is mildly over-dispersed relative to the ideal $\chi^{2}$, with a false rejection rate of $5.7\%$ at the nominal $5\%$ level in the toy. This is a small residual effect that does not affect the conclusions.

The method assumes that the event-sharing structure is exact. For this reason we recommend that analyzers stick to exact event-sharing structures when possible. Departures from this assumption are detectable through the spectral gap of the statistical covariance, the comparison of the empirical and structural ranks, and the consistency of the measured block totals, and we have given a corresponding set of diagnostics and remedies. Where the full forward model and nuisance parameterization are available, a binned likelihood fit handles the constraints automatically and is more robust to imperfect sharing; the range-projected $\chi^{2}$ instead requires only the reported cross-block covariance, the binning, and the detector response, making it suited to the goodness-of-fit testing, model tuning, and generator comparison that reuse published measurements.

As multi-distribution measurements with shared events become more common, a correctly calibrated global goodness-of-fit test becomes a routine requirement for their interpretation. The algorithm for constructing the structural combination matrix is given in Appendix~\ref{app:algorithm}, and a reference implementation can be found in Ref.~\cite{GlobalChi2Github}.

\acknowledgments

We thank Steve Pate-Morales, Andy Furmanski, Ben Bogart, and Nitish Nayak for valuable discussions that contributed to the development of this work, which originated from studies conducted within the MicroBooNE collaboration. P.~Green is supported by a United Kingdom Research and Innovation (UKRI) Future Leaders Fellowship, MR/V022407/1. This manuscript has been authored by Fermi Forward Discovery Group, LLC under Contract No.~89243024CSC000002 with the U.S. Department of Energy, Office of Science, Office of High Energy Physics.

\appendix
\section{Structural combination matrix algorithm}
\label{app:algorithm}

We describe the algorithm implemented to construct the structural combination matrix $\Omega_{\text{struct}}$ from the bin edge specifications of the multi-block measurement. Each row of $\Omega_{\text{struct}}$ is a unique indicator vector encoding one structurally allowed combination of bins, one per block, that a single event can simultaneously populate. The rank of this matrix determines the number of null directions, as described in Sec.~\ref{subsec:null_space}.

The construction proceeds in three stages. First, the kinematic variables used across all blocks are classified as shared (appearing in two or more blocks) or private (appearing in exactly one block). This distinction is key to the efficiency of the algorithm. Shared variables create correlations between blocks, while private variables contribute bins that are independent of all other blocks.

Second, a common refinement is computed for each shared variable by taking the union of all bin edges from every block that uses it. The algorithm then enumerates the cells of this shared-variable refinement. For each cell, the midpoint in each shared variable is used to identify which bin (or set of bins) in each block is compatible with that cell. When the binning of a given block is fully determined by the shared variables (e.g., a 1D block whose variable is shared, or a 2D block whose slice and bin variables are both shared) the cell maps to a single bin. When a block uses a private variable, the shared cell constrains only part of the binning, and the block contributes multiple candidate bins corresponding to the different private-variable choices. A 1D block whose variable is private contributes all of its bins; a 2D block with a shared slice variable and a private bin variable contributes all bins within the determined slice.

Third, the candidate bins from each block are combined. In the simplest case, the full Cartesian product of all candidate bins is expanded, and each element becomes a row of $\Omega_\text{struct}$. However, when many blocks have private variables, this product can become very large. In that case, a representative subset of indicator vectors is used instead. A base combination is formed by selecting the first candidate bin from each block, and then the remaining candidates from each block are substituted one at a time while the others are held at their base values. This produces at most $1 + \sum_{b} (n_{b} - 1)$ rows per shared cell, where $n_{b}$ is the number of candidate bins for block $b$. The representative set spans the same column space as the full Cartesian product, and therefore yields the same $\mathrm{rank}(\Omega_\text{struct})$ and null-space structure.

To understand why, note that any indicator vector from the full product can be written as the sum of indicator vectors from the representative set. Denoting the base combination as $\mathbf{e}_{0}$ and the single-block variations as $\mathbf{e}_j^{(k)}$ (block $k$ switched to its $j$-th candidate), any full-product row that selects candidate $j_k$ in each block satisfies
\begin{equation}
    \mathbf{e}_{j_1, \ldots, j_B} = \mathbf{e}_{0} + \sum_k (\mathbf{e}_{j_{k}}^{(k)} - \mathbf{e}_{0}),
\end{equation}
since the blocks occupy disjoint columns. The full product therefore lies in the span of the representative rows, and the rank is preserved.

When the measurement involves mutually exclusive event selections, blocks belonging to different selections share no events and are processed independently. The algorithm is applied separately to each selection group, and the resulting matrices are vertically concatenated.

The cost of the algorithm is dominated by the enumeration of shared-variable cells, which scales as $\prod_{v \in \text{shared}} M_v$ where $M_v$ is the number of refined bins for shared variable $v$. Since only shared variables enter this product, measurements with many blocks but few shared variables, such as a set of 1D distributions in different kinematic variables, are handled efficiently regardless of the total number of bins.


\bibliographystyle{JHEP}
\bibliography{biblio.bib}

\end{document}